\documentclass{aa}
\usepackage{txfonts}
\usepackage{graphicx}
\usepackage{fancyhdr}

\begin{document}

\title{Studying the highly bent spectra of FR\,II$-$type radio galaxies with the KDA\,EXT model}   
\author{El\.zbieta Kuligowska}
\offprints{E. Kuligowska \email elzbieta@oa.uj.edu.pl}
\institute{Astronomical Observatory of the Jagiellonian University, Orla 171 Cracow, Poland}
\date{Received 10 July 2017}

\abstract 
{The Kaiser,
Dennett-Thorpe \& Alexander (KDA) (1997)\,EXT model, that is, the extension of the KDA model of FR(Faranoff \& Riley)II$-$type source evolution, 
is applied and confronted with the observational data for selected FR\,II$-$type radio sources with significantly aged 
radio spectra.}
{A sample of FR\,II$-$type radio galaxies with radio spectra strongly bent at their highest frequencies
 is used for testing the usefulness of the KDA\,EXT model.}
{The dynamical evolution of FR\,II$-$type sources predicted with the KDA\,EXT model is briefly 
presented and discussed. The results are then compared to the ones obtained with 
the classical KDA approach, assuming the source's continuous injection and self-similarity.}
{The results and corresponding diagrams obtained for the eight sample sources indicate that the KDA\,EXT model predicts the observed radio 
spectra significantly better than the best spectral fit provided by the original KDA model.}
{}


\authorrunning{}
\maketitle

\section{Introduction}

The analytical models of FR\,II$-$type (Fanaroff \& Riley 1974) radio sources are based on the source's internal physics and characterise them in terms 
of dynamics, energetics, and luminosity evolution. These models originate from the "standard model" for double radio sources (Scheuer 1974, 
Blandford \& Rees 1974) in which FR\,II sources are composed of two opposite jets of plasma ejected from the vicinity of the supermassive black hole 
as a result of the accretion process. A travelling jet encounters the interstellar medium (ISM) of its host galaxy, the intergalactic medium (IGM), and finally 
the environment in the galaxy cluster, the intracluster medium (ICM). Its interaction with those media results in the formation of
the terminal bow shock and the interface between the shocked and unperturbed external material and the relativistic 
electrons reaccelerated in the so-called hot spot regions. 

The more sophisticated models for FR II-type sources such as KDA (Kaiser, Dennett-Thorpe \& Alexander 1997), BRW (Blundell, Rawlings \& 
Willott 1999) and MK (Manolakou \& Kirk 2002) differ from the standard model in the assumptions on the transport mechanisms of the new particles 
between the jets, hot spots, and lobes. While the KDA model assumes a constant injection index for the particles, in the BRW model this index varies between
the different energy regimes. However, the MK model again assumes a constant injection index, but also takes into account the reacceleration of the particles
in the turbulent processes occurring during their transport to the lobe. It is worth adding that Barai and Wiita (2006) performed detailed quantitative tests of three popular models describing the evolution of FR\,II radio
sources (KDA, BRW and MK) based on multidimensional Monte Carlo simulations. Their main aim was to compare the predictions of these models with
the observational data (samples constructed based on low-frequency radio surveys revised Third Cambridge Catalogue of Radio Sources (3CRR, Laing et al. 1983), 
revised 6C (6CE, Rawlings et al. 2001) and 7C Redshift Survey (7CRS, Willot et al. 2001). Barai and Wiita concluded that none of the
examined models produce fully satisfactory results (acceptable fits to all the properties of the samples). However, their statistical
test suggests that the KDA model still appears to provide better results than the Manolakou \& Kirk and Blundell et al. models.

The KDA\,EXT model originally published by Kuligowska (2017) arose from the need to better understand
the processes occurring in FR\,II$-$type radio sources after the end of their
nuclear activity. In contrast to the previously developed models describing the dynamics of these
sources and based on the "standard model", such as KDA, BRW and MK, KDA\,EXT was
constructed on the assumption of the lack of so-called continuous injection (hereafter CI) of new particles
into the lobes at the time at which the given source is observed. This implies that the radio lobes are no longer
powered by the central active galactic nucleus (AGN), and the whole source's structure begins to dissipate at radio wavelengths.

Non-thermal continuum radio emission of the lobes is due to the
synchrotron process and to the inverse-Compton scattering of ambient photons
of the cosmic microwave background. The synchrotron radiation results from the
ultra-relativistic particles interacting with the magnetic field, with the
most energetic ones emitting energy at high radio frequencies and losing it at the fastest rate.
Thus in the case of sources with no injection of new particles from their nucleus,
one should be able to observe a substantial steepening of their spectra. In fact such FR\,II-type
sources, with radio spectra highly bent at GHz frequencies, are known.
Among them are both very weak radio relics (Shulevski et al. 2015) and objects with so-called recurrent activity
(Saikia \& Jamrozy 2010), as well as the more typical, powerful radio galaxies with curved spectra (i.e., the ones
modeled and discussed by Kuligowska (2017). It is worth emphasising that the problem of evolution of FR\,II-type
sources after their jet activity has stopped was previously widely discussed by Kaiser \& Cotter (2002). They argue that, after the jet 
switches off, the information about ceasing of the lobes should propagate with the speed of sound.
If the internal sound speed in the lobes is relatively low, the adiabatic evolution of the cocoon is unchanged
for a long period of time. Lobes may still be overpressured with respect to the ambient gas and continue their expansion (model A).
However, if the sound speed is fast, the entire radio structure almost immediately transforms the so-called coasting phase and its
adiabatic evolution slows down rapidly (model B).

In the present paper, the suitability of the KDA\,EXT model for studying the FR\,II-type sources with bent spectra is
further analysed. However, instead of finding examples of such sources in a wide range of available literature or
radio catalogues, the present study is focused on the large and representative sample of 388 bright radio sources observed at 74 MHz
(Helmboldt 2008). This sample was used for the several reasons:  It is big enough to ensure the presence of sources
with various spectral shapes;  it consists of sources that are powerful at low radio frequencies, allowing reliable
modelling of their dynamics with KDA or KDA-like models; and it provides ready-to-use flux densities of the sources, available over
a wide range of the radio spectrum. Here, the analytical formulae for the dynamics and luminosity
evolution of the lobes, given by the KDA\,EXT model, is applied to the eight
 FR\,II-type sources with presumably terminated jet inflow carefully selected from the
Helmboldt 2008 sample.

The organisation of this article is as follows: the summary of the KDA\,EXT model and its comparison with the original KDA
model is provided in Section\,2. The description of the sample of Helmboldt (2008) is given in Section\,3 along with the
methodology for further selecting the sources with strongly curved spectra, for which the available observations
are sufficient to obtain satisfactory and accurate models. Section\,4 presents the method of fitting both the KDA\,EXT and KDA
models to the observational data of the selected radio galaxies, as well as the results (radio spectra obtained with the models).
Finally, Section\,5 presents a brief summary of these results.

\section{The summary of KDA and KDA\,EXT models}

\subsection{KDA and its background}

The KDA model is based on a dynamical description developed by Kaiser
\& Alexander (1997) and further combined with the radiative processes analysed by
Kaiser et al. (1997). Its most important assumptions are: power-law radial density distribution (King's 1972 model)
of unperturbed ambient gas surrounding the radio source, continuous injection of new particles into the
jet, and energy conservation within the jet. The model predicts (among others) the most evident observational parameters
of a radio source, that is, the total length of its jets (identified with lobes) and its total radio power at a given frequency. The
final equations for these parameters are given below, and their derivation is presented by Kaiser et al. (1997).

The length of the jet at a given source's age is given by

\begin{equation}
r_{\rm j}(t)=c_{1}\left(\frac{Q_{\rm j}}{\rho_{0}a_{0}^{\beta}}\right)
^{1/(5-\beta)}t^{3/(5-\beta)},
\end{equation}
where $\beta$ is the exponent of the density profile, $a_{0}$ is the size of the radio core,
$\rho_{0}$ is the central density of this core, $Q_{\rm j}$ is the jet’s power and $r_{\rm j}$
is identified with one-half
of the source’s linear size $D$, $r_{\rm j}=D/2$. If two of the model parameters,
$Q_{\rm j}$ and $\rho_{0}a_{0}^{\beta}$, are specified, the model
predicts the time evolution of the source's length.

The radio emission of the source is
calculated by splitting the cocoon surrounding the jet into infinitesimal evolving volume elements and
tracking its total adiabatic and radiative losses. The sum of the
contributions from volume elements gives the total emission (at a
frequency $\nu$), $P_{\nu}(t)$, as a integral over the injection
time $t_{\rm i}$, which can then be solved numerically.

\[
P_{\nu}(t) = 
\int\limits^t_{t_{\rm min}} \, dt_{\rm i} \frac{\sigma_{T}c\;r}{6{\pi}\nu(r+1)} 
Q_{\rm j}n_{0}(P_{\rm hc})^{(1-\Gamma_{\rm c})/\Gamma_{\rm c}}
\]
\vspace{-4mm}
\begin{equation}
 \times\frac{\gamma^{3-p}t_{\rm i}^{a_{1}/{3(p-2)}}} 
{[{t^{-a_{1}/3}}-a_{2}(t,t_{\rm i})\gamma)]^{2-p}} 
 \left(\frac{t}{t_{\rm i}} \right)^{-a_{1}(1/3+\Gamma_{\rm B})},
\end{equation}
\vspace{2mm}
\noindent  
where $(P_{\rm hc})$, the ratio of the jet head pressure $(p_{\rm h})$ and
the uniform cocoon pressure $(p_{\rm c})$, is a function of $(R_{\rm T})$ -
the axial ratio of the cocoon described by the empirical formula adopted
from Kaiser (2000), $\Gamma_{\rm c}$ and $\Gamma_{\rm B}$ are the adiabatic indices
in the equation of state of the cocoon material and the magnetic field, respectively, $\sigma_{T}$ is the 
Thompson cross-section, $c$ is the speed of light, $r$ is the ratio of the energy density of the magnetic field $u{_B}$, 
$a_{1}$ and $a_{2}$ are the physical constants defined by Kaiser (1997), $n_{0}$ is the number density of electrons (Heavens \& 
Drury 1988), $p$ is the exponent of the energy distribution of jet particles, and $\gamma$ is their Lorentz factor.

\subsection{KDA\,EXT model}

The idea of developing the KDA\,EXT model grew out of the need for an estimation of the age
of the source at which its nuclear activity starts to cease (hereafter referred as $t_{\rm br}$). One can assume that
stopping this activity ceases the inflow of new particles into the jet and eventually
reduces the radio brightness of the entire radio structure. For such sources, the KDA model may not be
reliable (even if it  can formally be fitted to the observed data), because its basic condition of
a continuous injection process is no longer fulfilled. The extended model should then consider both
the new parameter describing the source's dynamics ($t_{\rm br}$) and the overall evolution of the source after the jet stops.

The KDA\,EXT model (Kuligowska 2017) resulting from these theoretical considerations allows us to predict the lobe (jet)
length and radio power of the source at a given frequency. It is assumed that the termination of the energy supply propagates
from the AGN to the lobes with a relatively low sound speed (Kaiser et al. 2000) and, after switching
off the jets, the adiabatic evolution of the lobes of the large sources does not change significantly
for a long time. In particular, the lobes may still be overpressured with respect to the external gaseous
environment and therefore continue to expand behind a bow shock, as described by Kaiser \& Cotter (2002).
In the paper of Kuligowska 2017, the effective formula for lobe length, describing its
adiabatic evolution after termination of the activity and compared to the analogous equation given by the KDA model,
is based on the work of Kaiser \& Cotter and derived as:

\begin{equation}
D(t, t_{\rm br})= \left \{{
c_{1}\left(\frac{Q_{\rm j}}{\rho_{0}a^{\beta}_{0}}\right)
^{1/{(5-\beta)}}t^{{3/{(5-\beta)}}} \hspace{20pt} {\rm for\,}\,t < t_{\rm br}  \atop { 
D(t_{\rm br})\left(\frac{t}{t_{\rm br}}\right)
^{\frac{2(\Gamma_{\rm c}+1)}{\Gamma_{\rm c}(7+3\Gamma_{\rm c}-2\beta)}} \hspace{17pt} 
{\rm for\,}\,t \geq t_{\rm br}}}\right.
,\end{equation}
where
$D(t_{\rm br})$ is the cocoon length at the time of switching off the energy
supply. Hereafter the Kaiser \& Cotter's approach in the limiting
case when the internal sound speed is fast (their model B) is adopted.

The energy-loss process due to the synchrotron and inverse-Compton
scattering of the Cosmic Microwave Background (CMB) photons is characterised by the energy break in the energy spectrum comprised of
particles with different energies. In the case of a source supplied by a constant flow of particles,
the highest break frequency (for the pitch angle $\theta=90\degr$) is given by:

\begin{equation}
\nu_{\rm br,CI}=C_{1}B\gamma_{\rm br}^{2}=
\frac{C_{1}}{C_{2}^{2}}\frac{B}{(B^{2}+B_{\rm iC}^{2})^{2}t^{2}}
,\end{equation}
\noindent
which is assumed to be valid for $t\leq t_{\rm br}$, where
$B$ and $B_{\rm iC}$ are the magnetic field strengths related to
the synchrotron and inverse-Compton losses, and $C_{1}$ and $C_{2}$
are the physical constants defined by Pacholczyk (1970).
However, if the pitch angle of the particle distribution is
highly isotropic (Jaffe \& Perola 1973), the break frequency
valid for $t > t_{\rm br}$ is given by:

\begin{equation}
\nu_{\rm br,JP}=\frac {C_{1}}{C_{2}^2} \frac{B}{\{(2/3B^{2}+B_{\rm iC}^2)(t-t_{\rm br})\}^2}
,\end{equation}
where $C_{1}/C_{2}^{2}=2.51422\times 10^{12}$ and $B_{\rm iC}=0.318(1+z)^{2}$ nT.

\vspace{2mm} 
Thus the analytical formula for the
total radio power of a source at a given frequency is the simple sum of these two integrals over time $t$.
In the context of the KDA\,EXT model, the first of them is the total source power until the time
of the jet switch-off, and the second one represents its radio power emitted from $t_{\rm br}$
until the actual age of the observed source, $t$. The total power of a source is given by:

\begin{equation}
P_{\rm \nu}(t)= \left \{{
P_{\rm\nu}(t_{\rm min},t_{\rm br}) + P_{\rm\nu}(t_{\rm br},t)\hspace{25pt} {\rm for} \hspace{5pt}t_{\rm br} > t_{\rm min} } \atop {
P_{\rm\nu}(t_{\rm min},t)\hspace{84pt} {\rm for} \hspace{5pt}t_{\rm br}\leq t_{\rm min}}\right.
,\end{equation}

where the first top term is the integral given by Equation\,2 with the limit of integration $t_{\rm min}$ changed to $t_{\rm br}$,
($t_{\rm min}$, defined by Kaiser et al. (1997), is the minimum injection time for which the lobe material is still radiating at a
given frequency $\nu$), and the second top term corresponds to Equation\,9  of
Kuligowska (2017). The lower integral refers to the classical KDA integration given by Equation\,2.

\section{Selected sample of FR\,II-type sources with aged radio spectra}
\subsection{Helmboldt's sample of 388 bright radio sources at 74 MHz}
~
The KDA\,EXT model is expected to predict observational parameters (i.e., jet length, radio power value at given frequency) of
those FR\,II-type radio sources in which nuclear activity is terminated or suspended. It is assumed that the
continuum injection KDA model does not provide accurate fits of their observed radio spectra. While most of the known sources are 
active and can be fit with the KDA model, KDA\,EXT is also able to reproduce the curved spectra of the dying 
radio sources. However, it should be emphasised that KDA\,EXT model is still useful in the study of active sources with the CI process: it represents a
generalisation of the classical KDA model, and it can be easily reduced to the KDA model by setting one of its free parameters
($t_{\rm br}$) to zero.

One can assume that jet termination usually entails a rapid decrease of the source's radio flux density which is most significant at the
highest frequencies (depending on the value of $t_{\rm br}$: from about 5 GHz and up to 20$-$50 GHz or even more) - as previously 
mentioned in Section 1. In this work we select sources with extremely bent spectra rather
than simply the sources with very steep radio spectra.

In the previous paper of Kuligowska (2017) six examples of such sources were presented. This previous "test" sample for the KDA\,EXT model
was compiled from literature and composed of a variety of radio galaxies, including giant radio galaxies (GRGs) and radio sources with 
restarting activity. Although the final sample included some relatively young objects, it was constructed with the key assumption
that predominantly old and rather large sources should have their radio spectra significantly bent
towards the highest frequencies. Here this presupposition is abandoned. It is rather simply assumed that any sources with bent spectra may be
studied with the KDA\,EXT model, regardless of their other observational parameters, such as linear size, age, or redshift.
The additional condition to be met by such sources is their relatively high radio brightness - in particular, the high value of radio flux
density at low frequencies, which gives a better chance of good spectrum coverage over a wide range of radio frequencies (cf. Sect. 3.3). 

For the above reasons, the sample of 388 radio sources compiled by Helmboldt et al. (2008) should be perfect in the search for those
sources. First, it seems to be big enough to provide a representative subsample of objects of interest. Second, it was compiled
from the available literature for all sources within the VLA Low Frequency Sky Survey (VLSS) brighter than 15 Jy at 74 MHz, providing only the 
most powerful radio sources. Moreover, the sample also contains carefully collected radio frequency spectra for these sources, provided in
the literature and compiled using the corrections needed to put the flux densities from all
the references on the same absolute flux density scale of Baars et al. (1977). It follows that the sample of Helmboldt gives a convenient list
allowing research on the ready-made radio spectra of these sources. It is also worth noting that the flux densities measured by VLSS
are more reliable than many others from older low-frequency surveys, and the sky coverage of VLSS itself is large enough
to provide a satisfactory spatial diversity of the sample sources.

\subsection{Selection criteria}
~
As mentioned above, the primary criterion is the strong collapse of the source's radio spectrum at the highest frequencies. To extract such
sources, firstly all the source spectra provided by Helmboldt were visually inspected, and all objects without highly bent spectra, typical 
for sources with continuous activity, were rejected, as well as those which were
evidently not FR\,II-type radio sources, due to previous classification or their different spectral shapes (i.e.,
 compact steep spectrum (CSS) sources with spectra peaking around 100 MHz or lower, or GHz peaked spectrum (GPS) sources with inverted
spectra). After this initial step,
47 previously selected  radio sources were carefully examined for the completeness of the observational data required for the application
of the KDA\,EXT model. In particular, this model needs broad (wide enough to include very low and high frequencies) coverage of
the radio spectrum (this criterion was valid for all cases) and a redshift value determined for the parent radio galaxy (unknown for 
almost one third of the initially selected objects). However, the most 
important was the availability of the radio maps. They are not only necessary to correctly determine the length and thickness of the
lobes (input parameters for both KDA\,EXT and KDA models), but also to confirm the FR\,II-type morphology of the given source,
provided by the literature or online data catalogues. In particular, radio maps usually clearly show the presence of bright radio
lobes and hot spots.
Unfortunately, good-quality radio maps are not available in the case of 15 sources of the preliminary  sample. Moreover,
despite the shapes of their radio spectra, the maps of two previously selected sources (3C048 and NGC1692) do not show the
characteristic FR\,II morphology. After further verification, 3C048 turns out to be an example of a core-dominated source
(Laing et al. 1983), while NGC1692 (0453$-$20) was classified as a source with "diffuse and amorphous radio emission" by Morganti et al. (1993).
The radio galaxy 3C217 fulfils all the selection criteria, however it was previously examined with the same KDA\,EXT model in the
earlier paper by Kuligowska (2017). For this reason its KDA\,EXT analysis is not repeated in this work, and the reader is referred to that publication.

In the next step, the progression of the radio spectrum slope for the remaining 26 sources was further investigated.
The spectral index $\alpha_{\nu}$, changing along with the spectrum, was measured and its slope gradient was calculated for
every pair of neighbouring radio flux densities. Here the author assumes that the so-called high curvature (bend) of the given spectrum must result
in: 1) high ($\alpha_{\nu}\sim$ 1.5) spectral indices occurring in the highest-frequency part of the spectrum, implying a
high-frequency slope significantly exceeding the limiting value of $\alpha_{\rm inj}+0.5$ expected in the CI process;
2) violent steepening of the spectrum between the frequencies in the range of 1400 MHz up to 4800, 10000 MHz or even 20000 MHz, giving
an increment of $\alpha_{\nu}$ of nearly one between these data points; and 3) relatively low spectral index (not exceeding 0.6) for
the low-frequency flux densities (26$-$178 MHz).

The above considerations finally led to the further limitation of the initial sample to nine sources: 3C006.1, 3C032, 3C041, 3C123, 3C247,
3C300, 3C325, 3C401 and PKS0235-19. Further, 3C300 was excluded because its radio spectrum is very irregular and does not allow for a reliable fit of the KDA\,EXT model due to its repeated breaks and flattening at
higher frequencies. The final sample includes almost exclusively 3C sources and consists of objects of rather mid-scale linear sizes (68$-$386 kpc)
and relatively low (z up to 0.84) redshifts, except 3C325 whose
high (z$=$1.135) redshift makes this source especially interesting from the viewpoint of studying the processes of interruption of AGN activity.

In summary, the final number of dying FR\,II-type sources selected from Helmboldt's compilation and fulfilling the formal
criteria (specific steepening of their spectra) is of the order of 21, which is 5.4$\%$ of the initial sample. However, only 8 of these are
further investigated here for the reasons given above (lack of sufficient data, irregular spectrum).
At this point it would be interesting to provide the percentage of all the sources with terminated jet activity drawn from the complete population
of FR\,II- and FR\,I-type sources. Unfortunately, this is rather problematic due to the difficulty in estimating the total number of such
radio sources as well as detecting the weak radio emission of dying radio galaxies. However, Giovannini et al. (1988) argue that only a few
percent of all radio sources are dying, based on data from Bologna (B2, Colla et al. 1970) and Third Cambridge Catalogue of Radio Sources (3C, Edge et al. 1959) catalogues. This result is in line with the percentage obtained for 
Helmboldt's sample. It is also worth emphasising (Murgia et al. 2011) that dying radio sources are not synonymous with the radio relics, 
typically not associated with individual galaxies.

\begin{center}
\begin{table*}[t]
\caption{\small Observational data for FR II-type sources selected from the sample of Helmboldt (2008).}
\label{table:1}
\footnotesize
\begin{tabular}{@{}llllllllll}
\hline \hline
Name & & 3C006.1 & 3C32 & 3C41 & 3C123 & 3C247 & 3C325 & 3C401 & PKS 0235$-$19  \\
\hline
$z$ & & 0.840$\pm$\,0.0004 & 0.400 & 0.795$\pm$\,0.001 & 0.2177$\pm$\,0.0003 & 0.7489 & 1.135 & 0.201 & 0.620 \\
$LAS [\arcsec]$ & & 32$\pm$\,3 & 72$\pm$\,1 & 26$\pm$\,1 & 41$\pm$\,3 & 25$\pm$\,3 & 29$\pm$\,4 & 24$\pm$\,2 & 45$\pm$\,2 \\
$R_{t}$ & & 3.3$\pm$\,2.2 & 4.6$\pm$\,0.6 & 4.6$\pm$\,1.6 & 4.0$\pm$\,1.5 & 1.5$\pm$\,0.7 & 1.7$\pm$\,1.0 & 3.0$\pm$\,1.7 & 1.5$\pm$\,0.27  \\
$D [kpc]$ & & 246$\pm$\,23 & 386$\pm$\,5 & 200$\pm$\,8 & 123$\pm$\,11 & 188$\pm$\,22 & 235$\pm$\,33 & 68$\pm$\,7 & 310$\pm$\,14  \\
\hline
$S_{74} [Jy]$ & & 23.93\,$\pm$\,2.39 & 38.63\,$\pm$\,3.86 & 19.68\,$\pm$\,1.97 & 421.96\,$\pm$\,42.20 & 17.29\,$\pm$\,1.73 & 28.28\,$\pm$\,2.83 & 31.38\,$\pm$\,3.14 & 35.66\,$\pm$\,3.57\\
\hline
\hline
\end{tabular}
\end{table*}
\end{center}

\subsection{The summary of observational data for the sample sources}

Table~1 presents the observational data of the sample sources used for testing the usefulness of the KDA\,EXT model.
Redshifts are given with the highest known accuracy available in the literature. Despite the undefined values of some redshift 
uncertainties, it is assumed that the errors of the given spectroscopic redshifts are very small 
and do not affect the result of modelling. The `largest angular sizes' ($LAS$) and $R_{t}$ values of axial sources
are estimated based on the accessible radio maps (along with the uncertainties given by their angular 
resolution). The projected linear sizes of the sources, $D$,
are calculated following the formula $D=4.848\times10^{-6}L_{\rm A}(z)LAS$ and their measurement uncertainties are calculated
with the partial derivatives. Another model parameter, inclination angle of the jet axis $\theta$, is not listed in the table
because in the case of all the analysed sources it can be assumed to be close to 90 degrees (sources are not significantly
inclined to the line of sight). Moreover, it can be demonstrated that a slight deviation of this angle
does not significantly affect the modelling results.

Table~1 also includes the flux density values of the sources at 74 MHz. The remaining flux densities for a large set of frequencies and
their errors are derived from Helmboldt (2008). It is worth noting here that the original uncertainties on the measurements are
often underestimated (further discussed in the following Section). Due to the large number and variety of the available flux
densities, these data are not listed in the table, but are provided in the individual source summaries (Sect. 4).

Before fitting the model parameters (cf. following Section), all the flux densities derived following Helmboldt were inspected and some 
clearly undervalued data points 
(i.e. high frequency measurements covering the hot spots only) were removed from the input sets for the certain sources. Additionally,
the 150 MHz TIFR GMRT Sky Survey (TGSS, Intema et al. 2017) flux measurements available for all eight sources were added and used in the modelling.

The values of the sources' radio power $P_{\nu}$ are calculated from the given flux densities according to the formula
$P_{\nu}=S_{\nu} L_{\rm D}(z)^{2} (1+z)^{(\alpha_{\nu}-1)}$, where $L_{\rm D}(z)$ is the luminosity
distance of the given source determined assuming a flat Universe with Hubble constant $H_{0}$=71 ${\rm km \, s^{-1}Mpc^{-1}}$ 
and the Lambda cold dark matter ($\Lambda$CDM) model with cosmological
$\Omega_{m}$=0.27 and $\Omega _{\Lambda}$=0.73, and where $\alpha_{\nu}$ is the spectral index measured as the
spectrum slope gradient calculated separately for every pair of neighbouring radio flux densities.
Moreover, determining the initial dynamical parameters for both the KDA and KDA\,EXT models requires at least five sources' radio power
$P_{\nu}$ values at significantly distant
observing frequencies for the reliable application of the models. 

We include recent 150 MHz TGSS flux densities to increase the number of measurements and accuracy of the calculations.
However, for the sources 3C41, 3C123, and PKS 0235-19 the 150 MHz data points clearly stand out from the trend lines. In these cases, the use of the
TGSS flux densities in the modelling does not improve the accuracy of the fits, and the use of these values is debatable. Moreover,
the TGSS flux densities of some sources are recognised as having systematically underestimated values (up to even 50\%). Special fixes to the
original TGSS, resulting in the revised Rescaled Subset of the Alternative Data Release 1 (TGSS-RSADR1) catalogue, were recently proposed by Hurley-Walker (2017). Thus TGSS flux densities
for these three sources have been verified by cross checking bright point-like sources in their vicinity (comparing their TGSS
flux densities with the flux densities observed at close frequencies). For 3C41, both the target and the sources in its close neighbourhood have
their TGSS flux densities a few percent lower than the values obtained in the 6CVI survey at 151 MHz. In the case of 3C123 and PKS 0235-19 the
situation is more complex. Their TGSS flux densities (and those of the sources in their vicinity) are a few percent larger than
the corresponding flux densities derived from the GaLactic and Extragalactic All-sky MWA survey (GLEAM, Hurley-Walker et al. 2017) at 150.5 MHz.
Also, the TGSS-RSADR1 catalogue is not fully conclusive here: for 3C123, it provides a flux density value even lower than the one obtained
with the original TGSS (271.70 vs. 272.08 Jy), while in the case of PKS 0235-19 the flux is actually larger and has a value of 24.20 Jy (vs. 24.12 Jy for
TGSS). In the case of 3C41 there is no available flux density provided by TGSS-RSADR1 due to its limited range of declination. Nevertheless it can be
clearly proven that these small differences in the flux densities at 151 MHz do not significantly affect the results of the modelling and the final
statistical tests used to examine the compared models (cf. Sect. 4). In fact, changing these TGSS values to the values given by TGSS-RSADR1
only results in a difference in the third decimal place for the calculated likelihood ratios.

\section{Application of the KDA\,EXT model to the sample sources and its comparison to the KDA fits.}
\subsection{Fitting procedure}

In the theoretical considerations, the only unknown parameters are the source's physical parameters, such as injection 
index $\alpha_{\rm inj}$, age $t$, the jet's power $Q_{\rm jet}$ , and central density $\rho_{0}$. However, the original KDA model 
only enables the prediction of the time evolution of the source's observational parameters (jet length $D$, radio luminosity 
$P_{\nu}$ at a given frequency, the shape of radio spectrum, and the lobe's volume $V$) with the previous assumption about the 
intrinsic dynamics of this source, defined by the physical quantities mentioned above. This implies that it is first necessary to 
solve the reverse problem of estimating the initial values of $t$, $\alpha_{\rm inj}$, $Q_{\rm jet}$, and $\rho_{0}$ in order to 
subsequently fit the proper KDA or KDA\,EXT model to the observed source parameters. This was performed using the DYNAGE algorithm 
(Machalski et al. 2007). In this step, the age resulting from the KDA model (t$_{\rm KDA}$) for the low-frequency parts of the spectra of 
the sample sources was determined. Here the high-frequency flux densities mostly contributing to the high curvatures of the observed source 
spectra at above 2000 MHz were excluded from the input data. Additionally, the appropriate correction for the changed length of the jets was taken into 
account in the KDA model, neglecting the input flux densities at these high frequencies results in fitting the younger age of 
the source and the lower value of the length of its lobes). 
This has provided the estimation of the approximate values of four model 
parameters: $\alpha_{\rm inj}$, the source's dynamical age, $t$, jet power, $Q_{j}$; and the central core density, $\rho_{0}$. 
The remaining model parameters were also adopted as in Kuligowska (2007), for example, $\Gamma_{\rm c}$=$\Gamma_{\rm a}$=$\Gamma_{\rm B}$=$5/3$ 
for the "cold" equation of state, and $\beta$=1.5.

Determining the so-called “best age solution” for the initial KDA fit demanded testing some model parameters: 
here it was assumed that the source's age, $t$, may vary from $10^4$ up to $10^8$ years, while the $\alpha_{\rm inj}$ is from the rather 
wide range of 0.3 up to 1.0, and is fitted with the accuracy of 0.01. In the first (fitting the KDA values) and the second (determining $t_{br}<t$) steps,
the best age solutions were fitted with an accuracy of 0.01 million years (Myr). It should be noted that $t_{br}<t$ solutions are unambiguous: there are 
clear and single minima of the obtained $\chi^{2}$ values in the function of tested $t_{br}<t$ values.

In the second step, the values of source power $P_{\nu_{em}}$ given by Equation 6 at the given frequencies for 
a number of arbitrarily chosen values for the jet switch-off time $t_{br}$, fulfilling the $t_{br}<t$ condition, were calculated.  
Then, the best fit of these model results was determined using the least-squares method by minimizing the expression: 

\begin{eqnarray}
\chi^{2}_{red}=\frac{1}{n-r-1}\sum\limits_{n}\left(\frac{S_{\nu_{0}}-S_{MOD}}{\Delta S_{\nu_{0}}}\right)^2,
\end{eqnarray}
\noindent
where n is the number of the compared values (flux densities), r is the number of the model's free parameters, 
 $S_{\nu_{0}}$ and $\Delta S_{\nu_{0}}$ are flux densities, and $S_{MOD}$ are the model flux densities re-calculated from the model
values of the emitted power $P_{\nu_{em}}$ according to

\begin{eqnarray}
S_{\rm MOD}=P_{\nu_{em}} \left(\frac{1+z}{L_{D}^{2}(z)}\right) = P_{\nu_{0}(1+z)} \left(\frac{1+z}{L_{D}^{2}(z)}\right).
\end{eqnarray}

The $S_{\rm MOD}$ flux densities, resulting from the KDA (best age solution) and KDA\,EXT models, are given in Tables~2$-$9 along with the values of 
$\chi^{2}$, determining the goodness of the resulting fits, and the numbers of their free parameters, r. 
The accuracy of both the fitted models is compared to the observed flux densities representing the real radio spectra of the sample sources.
In addition, the results of the likelihood ratio test ($\Lambda$) were calculated for the compared KDA and KDA\,EXT best model fits, 
based on their $\chi^{2}$ distributions and for the assumed level of significance of 10$\%$ (0.01), in order to obtain the reliable 
statistical tool for the model selection (providing the formal statistical preference of one model over another). The use of this tool is justified here by the fact that the KDA and the KDA\,EXT models are nested (KDA\,EXT is a more general 
case of the KDA, meaning all the parameters of the simpler KDA model also occur in the KDA\,EXT, and the KDA model may be derived directly from KDA\,EXT by setting one of its free parameters, $t_{br}$, to 0). It is worth noting that the KDA\,EXT has only one more degree of freedom than the KDA (associated with $t_{br}$).

The formal values of the flux density uncertainties resulting from the KDA and KDA\,EXT models are also provided. These uncertainties include both the 
uncertainties of the calculated radio powers ($P_{\rm \nu}$), and the errors of the redshift values known from the literature. 
It should be noted, however, that these uncertainties are rather crude, especially due to the lack of reliable redshift uncertainties. They are known 
only for three of the eight analysed sources (cf. Table~1), and for the remaining ones the rough values of the redshift error are equal 
to 0.001 (the higher known redshift uncertainty from the examined sources sample) have been adopted. As a result, the final flux density errors 
may be overvalued or underestimated. Such approximate uncertainties, however, are only supposed to provide an overall view on the capabilities 
and limitations of the KDA and KDA\,EXT models (cf. Sect. 5). 

\subsection{Fitting results}
~
Table~10 presents the values of the model free parameters and derived physical
parameters of the sample sources resulting from the best fit of the KDA and KDA\,EXT models. These
derived parameters are the cocoon pressure, $p_{\rm c}$; the total emitted energy,
U$_{\rm c}$; the strength of the magnetic field, B$_{\rm eq}$; and the radial expansion speed
of the cocoon's head, v$_{\rm h}$. It is worth noting that their values are calculated
for t $>$ t$_{br}$ (after the switch-off), analogous to the formulae given by Kuligowska (2017). Namely, 
v$_{\rm h}(t)$ and $p_{\rm c}$ for t>t$_{br}$ are calculated with Equations 11 and 12 of Kuligowska 2017, respectively, 
and U$_{\rm c}$ is given by the formula U$_{\rm c}(t)$=u$_{\rm c}(t)$V$_{\rm c}(t)$=p$_{\rm c}(t)$V$_{\rm c}(t)/(\Gamma_{c}-1)$.
A compilation of the results obtained for the individual sources from the sample is presented below.
In this paper the given errors have been enlarged due to their initial incompleteness, presumably resulting from not taking into account
all of the error components. Following the typical percentage error values estimated, for example, for VLSS flux
densities, it was arbitrarily assumed that the actual error values can be up to 10\%\ of the given flux density
for the observational frequency range 20$-$750 MHz, and equal to about 5\% for the higher frequencies.
The $\chi^{2}$ values in Tables$~$2-9 are derived with the errors defined in the above way.

\subsubsection{3C006.1}

~3C006.1 is an example of classical FR\,II-type radio source with long lobes and
distinct hot spots. It lies at relatively high redshift (z$=$0.84) and was previously
studied by Neff et al. (1995). It is worth noting that the radio core of this galaxy was not
detected in the VLA maps at 6 cm, and the authors argue that its nucleus may have changed its spatial orientation.
This guess is based on radio morphology of the source: a faint inner lobe structure is visible on VLA radio maps at 20 cm.

The radio spectrum coverage for this source extends from 74 {\rm MHz} up to 21.7 {\rm GHz}. Such a broad spectrum
usually results in reliable KDA\,EXT or KDA model fits to the observed flux densities.
However, in this case the strong bend typical for old FR\,II-type sources is visible only at
extremely high frequencies (starting from 10.7 GHz), and both fitted models are rather poor, especially
in comparison with other objects from the sample. This source also has a very flat low-frequency
part of the spectrum (spectral index between 74 and 178 MHz equal to 0.45), making it difficult to determine 
the initial parameters of the models. As a result, spectra obtained from the predicted flux densities $S_{\rm MOD}\,(KDA)$ and 
$S_{\rm MOD}\,(KDA\,EXT)$ (Fig. 1) depart from both low and high-frequency data points, though the KDA\,EXT visually 
fits better to the observed flux densities and the provided result for the likelihood ratio test clearly indicates a preference for the 
KDA\,EXT over the KDA model.

\begin{table}[h]
\caption{\small Model flux densities $S_{MOD}$ from the KDA and KDA\,EXT fitting for 3C006.1 and their goodness of fit to the observed data $S_{\nu_{0}}$.}
\footnotesize
\begin{center}
\begin{tabular}{@{}rrrr}
\hline
$\nu_{0}$ [MHz] & $S_{\nu_{0}}\pm\Delta$ [Jy] & $S_{MOD}$ [Jy] & $S_{MOD}$ [Jy] \\
         & & KDA\,EXT  & KDA\\
\hline
\hline
74 &     23.93\,$\pm$\,2.39 &           32.32\,$\pm$\,3.24 &     31.44\,$\pm$\,3.24 \\
150 &    \,\,\,18.27\,$\pm$\,1.83  &    19.21\,$\pm$\,1.93  &    18.48\,$\pm$\,1.93 \\
178 &    \,\,\,15.13\,$\pm$\,1.51 &     16.84\,$\pm$\,1.69  &    16.36\,$\pm$\,1.69 \\
232 &    \,\,\,13.45\,$\pm$\,1.34 &     13.69\,$\pm$\,1.37  &    13.23\,$\pm$\,1.37 \\
325 &    9.96\,$\pm$\,1.00 &            10.46\,$\pm$\,1.05 &     9.98\,$\pm$\,1.05 \\
1400 &   3.65\,$\pm$\,0.36 &            3.06\,$\pm$\,0.30  &     3.00\,$\pm$\,0.30 \\
2260 &   2.20\,$\pm$\,0.22 &            2.01\,$\pm$\,0.20  &     1.97\,$\pm$\,0.20 \\
2695 &   \,\,\,1.87\,$\pm$\,0.19 &      1.73\,$\pm$\,0.17  &     1.68\,$\pm$\,0.17 \\
3900 &   \,\,\,1.42\,$\pm$\,0.14 &      1.25\,$\pm$\,0.13  &     1.20\,$\pm$\,0.13 \\
4800 &   \,\,\,1.10\,$\pm$\,0.06 &      1.03\,$\pm$\,0.10  &     1.00\,$\pm$\,0.10 \\
4900 &   \,\,\,1.07\,$\pm$\,0.05 &      1.02\,$\pm$\,0.10  &     0.98\,$\pm$\,0.10 \\
5000 &   1.03\,$\pm$\,0.05 &            0.99\,$\pm$\,0.10  &     0.96\,$\pm$\,0.10 \\
7700 &   0.64\,$\pm$\,0.03 &            0.66\,$\pm$\,0.07  &     0.65\,$\pm$\,0.07 \\
8390 &   0.60\,$\pm$\,0.03 &            0.61\,$\pm$\,0.06  &     0.60\,$\pm$\,0.06 \\
10700 &  \,\,\,0.46\,$\pm$\,0.02 &      0.47\,$\pm$\,0.05  &     0.48\,$\pm$\,0.05 \\
11200 &  \,\,\,0.42\,$\pm$\,0.02 &      0.45\,$\pm$\,0.05  &     0.46\,$\pm$\,0.05 \\
14900 &  \,\,\,0.27\,$\pm$\,0.01 &      0.31\,$\pm$\,0.03  &     0.35\,$\pm$\,0.03 \\
21740 &  \,\,\,0.14\,$\pm$\,0.01 &      0.18\,$\pm$\,0.02  &     0.25\,$\pm$\,0.02 \\
\hline
$r$ &  & 4 & 3 \\
$\chi^{2}$ &  & 4.97 & 21.39 \\
& $\Lambda$=0.03 & ($p$=0.1) &   \\
\hline
\end{tabular}
\end{center}
\end{table}

\subsubsection{3C32}

The radio source 3C32 is the largest source in the examined sample, with its lobes totalling 385.7 kpc in length. Its central AGN 
have been classified as Seyfert type 2 galaxy by Véron-Cetty \& Véron (2006).  
The spectra resulting from the predicted flux densities $S_{\rm MOD}\,(KDA)$ and
$S_{\rm MOD}\,(KDA\,EXT)$ are shown in Figure 2. In this case, the preference of the KDA\,EXT model over KDA model 
is evident, with the $\Lambda$ value of 0.002 for a level of significance of 10$\%$. However, the goodness of the fit to data points ($\chi^{2}$ is not high and some observational 
data points are very far away from the trend line (compare Table~3)).

\begin{table}[h]
\caption{\small Model flux densities $S_{MOD}$ from the KDA and KDA\,EXT fitting for 3C32 and their goodness of fit to the observed data $S_{\nu_{0}}$.}
\footnotesize
\begin{center}
\begin{tabular}{@{}rrrr}
\hline
$\nu_{0}$ [MHz] & $S_{\nu_{0}}\pm\Delta$ [Jy] & $S_{MOD}$ [Jy] & $S_{MOD}$ [Jy] \\
         & & KDA\,EXT  & KDA\\
\hline
\hline
74      &       38.63   $\pm$   3.86    &       44.37 $\pm$ 4.47                &       51.89 $\pm$ 5.23      \\
150     &       27.82   $\pm$   2.78    &       27.30 $\pm$ 2.75                &       27.82 $\pm$ 2.80      \\
160     &       23.60   $\pm$   2.36    &       25.59 $\pm$ 2.58                &       26.43 $\pm$ 2.66      \\
365     &       14.23   $\pm$   1.42    &       13.40 $\pm$ 1.35                &       12.36 $\pm$ 1.25      \\
750     &       8.59    $\pm$   0.86    &       7.28 $\pm$ 0.73                 &       6.20 $\pm$ 0.62      \\
1400    &       4.20    $\pm$   0.21    &       4.14 $\pm$ 0.42                 &       3.36 $\pm$ 0.34      \\
2695    &       2.11    $\pm$   0.11    &       2.19 $\pm$ 0.22                 &       1.74 $\pm$ 0.18      \\
2700    &       2.25    $\pm$   0.11    &       2.19 $\pm$ 0.22                 &       1.74 $\pm$ 0.18      \\
4850    &       1.23    $\pm$   0.06    &       1.18 $\pm$ 0.12                 &       0.96 $\pm$ 0.10      \\
5009    &       1.17    $\pm$   0.06    &       1.14 $\pm$ 0.11                & 0.93 $\pm$ 0.09 \\
10695   &       0.31    $\pm$   0.02    &       0.43 $\pm$ 0.05                 &       0.43 $\pm$ 0.04      \\
\hline
$r$ &  & 4 & 3 \\
$\chi^{2}$ &  & 9.80 & 23.37 \\
&  $\Lambda$=0.002 & ($p$=0.1) &   \\
\hline
\end{tabular}
\end{center}
\end{table}

\subsubsection{3C41}

Radio source 3C41 shows classic extended FR II-type morphology and lies at relatively high $z$. 
It is worth noting that its KDA\,EXT model fit, presented on Figure 3, seems very reliable, with a resulting 
$\chi^{2}$ value almost ten times less than the corresponding value for KDA model. The 
likelihood ratio test, however, suggests that there is no significant preference of the KDA\,EXT model over KDA, or 
at least the results of both models are statistically comparable, with $\Lambda$ slightly higher than the assumed 
significance level.
The spectra calculated from both fitted models are also given in Table~4. 
Interestingly, in the case of 3C41, the flat low-frequency part of the KDA\,EXT radio
spectrum  visually fits the observed data well, regardless of the poor statistical $\Lambda$ result.

\begin{table}[h]
\caption{\small Model flux densities $S_{MOD}$ from the KDA and KDA\,EXT fitting for 3C41 and their goodness of fit to the observed data $S_{\nu_{0}}$.}
\footnotesize
\begin{center}
\begin{tabular}{@{}rrrr}
\hline
$\nu_{0}$ [MHz] & $S_{\nu_{0}}\pm\Delta$ [Jy] & $S_{MOD}$ [Jy]  & $S_{MOD}$ [Jy]\\
         & & KDA\,EXT  & KDA\\
\hline
\hline
74 &    19.68   $\pm$   1.97    &       21.47 $\pm$ 2.98        &       29.71 $\pm$ 2.16  \\
112 &   15.51   $\pm$   1.55    &       17.33 $\pm$ 2.25        &       22.33 $\pm$ 1.74  \\
150 &   12.20   $\pm$   1.22    &       14.75 $\pm$ 1.82        &       18.11 $\pm$ 1.48  \\
160 &   14.20   $\pm$   1.42    &       14.32 $\pm$ 1.74        &       17.29 $\pm$ 1.44  \\
232 &   11.21   $\pm$   1.12    &       11.69 $\pm$ 1.32        &       13.15 $\pm$ 1.18  \\
325 &   9.39    $\pm$   0.94    &       9.63 $\pm$ 1.03     &           10.21 $\pm$ 0.98  \\
365 &   9.40    $\pm$   0.94    &       9.01 $\pm$ 0.94         &       9.33 $\pm$ 0.91  \\
1400 &  3.71    $\pm$   0.19    &       3.82 $\pm$ 0.32        &        3.22 $\pm$ 0.39  \\
1410 &  3.65    $\pm$   0.18    &       3.80 $\pm$ 0.32        &        3.21 $\pm$ 0.38  \\
1670 &  3.43    $\pm$   0.17    &       3.37 $\pm$ 0.28         &       2.79 $\pm$ 0.34  \\
2695 &  2.26    $\pm$   0.11    &       2.36 $\pm$ 0.19   &     1.88 $\pm$ 0.24  \\
2700 &  2.30    $\pm$   0.12    &       2.35 $\pm$ 0.19  &      1.88 $\pm$ 0.24  \\
4850 &  1.47    $\pm$   0.07    &       1.47 $\pm$ 0.12  &      1.16 $\pm$ 0.15  \\
5000 &  1.46    $\pm$   0.07    &       1.44 $\pm$ 0.11  &      1.12 $\pm$ 0.14  \\
10695 & 0.76    $\pm$   0.04    &       0.73 $\pm$ 0.06 &       0.59 $\pm$ 0.07  \\
14900 & 0.51    $\pm$   0.03    &       0.53 $\pm$ 0.05 &       0.45 $\pm$ 0.05  \\
31400 & 0.21    $\pm$   0.01    &       0.24 $\pm$ 0.02 &       0.24 $\pm$ 0.02  \\
\hline
$r$ &  & 4 & 3 \\
$\chi^{2}$ &  & 1.62 & 15.37 \\
&  $\Lambda$=0.11 & ($p$=0.1) &   \\
\hline
\end{tabular}
\end{center}
\end{table}

\subsubsection{3C123}

This next radio source from the 3C catalogue is classified as a FR II-type galaxy with extended tails of both lobe components
(Fanaroff \& Riley 1974). It is also characterised by a relatively low redshift (z$=$0.218, Spinrad et al. 1985) and short lobes
(compare Table~1). 3C123 also shows exceptionally high radio luminosity and non-typical radio structure (Riley \& Pooley 1978). Its
hot spots along with the general spectral analysis were studied in detail by Looney \& Hardcastle (2000) with the highest resolution to
date (0.6 $\arcsec$). The authors concluded that 3C123 may have two pairs of (primary and secondary) hot spots, and there is no clear evidence for
a continuing outflow between these hot spots, suggesting that the source may be undergoing recurrent activity. If this source is 
actually restarting, one could argue it would be more effective
to use the model of Brocksopp et al. (2011) which also introduces a free parameter related to the age at which energy supply to the jet stops,
t$_{\rm j}$. Nevertheless, 3C123 meets the strict selection criteria described in Section 3.2 and here it is simply
assumed that the application of the KDA\,EXT model to the case of this source is also both justified and interesting.

The radio spectrum of this source is very broad thanks to its high radio brightness. Surprisingly, analysis shows that there is
no major difference between the spectra resulting from the predicted flux densities for KDA\,EXT and KDA models (Fig. 4), though
the KDA\,EXT fit is slightly more consistent with observations. There is also no statistical preference of the KDA\,EXT model,
according to the obtained $\Lambda$ value for the adopted level of significance (Table~5).

\begin{table}[h]
\caption{\small Model flux densities $S_{MOD}$ from the KDA and KDA\,EXT fitting for 3C123 and their goodness of fit to the observed data $S_{\nu_{0}}$.}
\footnotesize
\begin{center}
\begin{tabular}{@{}rrrr}
\hline
$\nu_{0}$ [MHz] & $S_{\nu_{0}}\pm\Delta$ [Jy] & $S_{MOD}$ [Jy] & $S_{MOD}$ [Jy] \\
         & & KDA\,EXT  & KDA\\
\hline
\hline
26      &       840.10  $\pm$   84.00   &       981.23 $\pm$ 98.38      &     1020.40 $\pm$ 102.31        \\
38      &       628.26  $\pm$   62.83   &       882.61 $\pm$ 88.46      &       913.63 $\pm$ 91.61     \\
74      &       421.96  $\pm$   42.20   &       691.46 $\pm$ 69.33      &       707.57 $\pm$ 70.95             \\
80      &       342.00  $\pm$   34.20   &       443.37 $\pm$ 44.46      &       446.45 $\pm$ 44.76             \\
150     &       272.08  $\pm$   27.21   &       274.26 $\pm$ 27.52      &       271.14 $\pm$ 27.12             \\ 
160     &       226.60  $\pm$   22.66   &       256.90 $\pm$ 25.78      &       255.13 $\pm$ 25.58             \\
178     &       208.25  $\pm$   20.83   &       237.56 $\pm$ 23.82      &       235.92 $\pm$ 23.66             \\
318     &       135.45  $\pm$   13.55   &       153.38 $\pm$ 15.37      &       150.93 $\pm$ 15.13     \\
325     &       151.07  $\pm$   15.11   &       150.58 $\pm$ 15.09      &       148.17 $\pm$ 14.86             \\
327     &       148.88  $\pm$   14.89   &       149.89 $\pm$ 15.01      &       147.49 $\pm$ 14.79             \\
408     &       122.34  $\pm$   12.23   &       126.12 $\pm$ 12.65      &       123.53 $\pm$ 12.39             \\
750     &  80.10 $\pm$  8.01    &       76.87  $\pm$ 7.71       &       74.95 $\pm$ 7.51              \\
1400    &   49.73 $\pm$ 2.49    &       45.16  $\pm$ 4.53       &       43.63 $\pm$ 4.37              \\
2695    &   27.50 $\pm$ 1.38    &       25.16  $\pm$ 2.052      &       24.14 $\pm$ 2.37              \\
2700    &      23.35 $\pm$      1.17    &       25.16  $\pm$ 2.052      &       24.09 $\pm$ 2.42              \\
4850    &      16.03 $\pm$      0.80    &       14.65  $\pm$ 1.48       &       13.89 $\pm$ 1.40              \\
8420    &      10.15 $\pm$      0.51    &       8.67 $\pm$ 0.87         &       8.12 $\pm$ 0.81              \\
10500   &      7.90  $\pm$      0.40    &       7.01 $\pm$ 0.70         &       6.53 $\pm$ 0.65              \\
10695   &      8.18  $\pm$      0.41    &       6.88 $\pm$ 0.69         &       6.41 $\pm$ 0.64              \\
14900   &      5.30  $\pm$      0.27    &       4.99 $\pm$ 0.50         &       4.60 $\pm$ 0.46              \\
15300   &      5.41  $\pm$      0.27    &       4.85 $\pm$ 0.49         &       4.47 $\pm$ 0.45              \\
22000   &      3.45  $\pm$      0.17    &       3.39 $\pm$ 0.34         &       3.10 $\pm$ 0.31              \\
\hline
$r$ &  & 4 & 3 \\
$\chi^{2}$ &  & 2.96 & 9.72 \\
& $\Lambda$=0.72 & ($p$=0.1) &   \\
\hline
\end{tabular}
\end{center}
\end{table}

\subsubsection{3C247}

3C247 is another example of a FR II-type radio galaxy at rather high redshift. It was previously studied by
Laing et al. (1983) who suggested that its parent galaxy coincides with a weak radio core. Its Multi Element Radio Linked Interferometer Network (MERLIN) radio map (Chidi et al. 1994)
shows rather thin and elongated lobes with bright hot spots.
The spectra resulting from the predicted flux densities for KDA\,EXT and KDA models are presented in Figure 5,
and the exact flux density values, along with the goodness of the fit to the data and the likelihood ratio analysis for the models,
are listed in Table~6. Although visually the KDA\,EXT model predicts the observed spectra of 3C247 with better accuracy, the resulting
$\Lambda$ value does not confirm a preference for this model over the KDA model.

\begin{table}[h]
\caption{\small Model flux densities $S_{MOD}$ from the KDA and KDA\,EXT fitting for 3C247 and their goodness of fit to the observed data $S_{\nu_{0}}$.}
\footnotesize
\begin{center}
\begin{tabular}{@{}rrrr}
\hline
$\nu_{0}$ [MHz] & $S_{\nu_{0}}\pm\Delta$ [Jy] & $S_{MOD}$ [Jy] & $S_{MOD}$ [Jy] \\
         & & KDA\,EXT  & KDA\\
\hline
\hline
74      &       17.29   $\pm$   1.73    &       18.64 $\pm$ 1.87        &       23.74 $\pm$ 2.39      \\
82      &       16.50   $\pm$   1.65    &       17.72 $\pm$ 1.78        &       22.06 $\pm$ 2.22      \\
150     &       13.64   $\pm$   1.36    &       13.02 $\pm$ 1.31        &       14.08 $\pm$ 1.42      \\
232     &       10.31   $\pm$   1.03    &       10.22 $\pm$ 1.02        &       10.11 $\pm$ 1.02      \\
325     &       8.40    $\pm$   0.84    &       8.37 $\pm$ 0.84         &       7.74 $\pm$ 0.78 \\
327     &       8.63    $\pm$   0.86    &       8.33 $\pm$ 0.84         &       7.70 $\pm$ 0.77      \\
365     &       7.92    $\pm$   0.79    &       7.77 $\pm$ 0.78         &       7.04 $\pm$ 0.71      \\
408     &       7.01    $\pm$   0.70    &       7.25 $\pm$ 0.73         &       6.44 $\pm$ 0.65      \\
750     &       4.93    $\pm$   0.49    &       4.81 $\pm$ 0.48         &       3.89 $\pm$ 0.39      \\
1400    &       2.88    $\pm$   0.14    &       3.02 $\pm$ 0.30         &       2.28 $\pm$ 0.23      \\
2695    &       1.60    $\pm$   0.08    &       1.75 $\pm$ 0.18         &       1.29 $\pm$ 0.13      \\
2700    &       1.66    $\pm$   0.08    &       1.75 $\pm$ 0.18         &       1.29 $\pm$ 0.13      \\
4800    &       0.93    $\pm$   0.05    &       1.02 $\pm$ 0.10         &       0.77 $\pm$ 0.08      \\
4850    &       1.01    $\pm$   0.05    &       1.01 $\pm$ 0.10         &       0.77 $\pm$ 0.08      \\
4900    &       0.96    $\pm$   0.05    &       1.00 $\pm$ 0.10         &       0.76 $\pm$ 0.08      \\
5000    &       0.940   $\pm$   0.05    &       0.98 $\pm$ 0.09        &        0.74 $\pm$ 0.07      \\
10695   &       0.400   $\pm$   0.02    &       0.42 $\pm$ 0.04         &       0.38 $\pm$ 0.04      \\
14900   &       0.250   $\pm$   0.01    &       0.28 $\pm$ 0.03         &       0.28 $\pm$ 0.03      \\
\hline
$r$ &  & 4 & 3 \\
$\chi^{2}$ &  & 1.35 & 11.63 \\
 & $\Lambda$=0.34 & ($p$=0.1) &   \\
\hline
\end{tabular}
\end{center}
\end{table}

\subsubsection{3C325}

With $z=$1.135, 3C325 is the most redshifted galaxy in the sample. It is characterised by a highly curved radio spectrum and a large
radio structure with bright lobes. Its classical FR II-type morphology is associated with a parent quasar (Véron-Cetty \& Véron 2006).
Preference of the KDA\,EXT over KDA solution is not evident: the calculated $\Lambda$ value is exactly equal to the assumed 
significance level for the adopted statistical test.
Figure 6 shows that this model fits the high-frequency part of the radio spectrum especially well, while departing from the low 
flux density data points except for the 150 MHz TGSS flux (see also Table~7).

\begin{table}[h]
\caption{\small Model flux densities $S_{MOD}$ from the KDA and KDA\,EXT fitting for 3C325 and their goodness of fit to the observed data $S_{\nu_{0}}$.}
\footnotesize
\begin{center}
\begin{tabular}{@{}rrrr}
\hline
$\nu_{0}$ [MHz] & $S_{\nu_{0}}\pm\Delta$ [Jy] & $S_{MOD}$ [Jy] & $S_{MOD}$ [Jy] \\
         & & KDA\,EXT  & KDA\\
\hline
\hline
74      &       28.28   $\pm$   2.83    &       37.64 $\pm$ 3.78        &       39.22 $\pm$ 0.65      \\
150     &       24.16   $\pm$   2.42    &       22.73 $\pm$ 2.28        &       21.88 $\pm$ 0.56      \\
178     &       15.85   $\pm$   1.59    &       20.07 $\pm$ 2.02        &       18.86 $\pm$ 0.45      \\
232     &       14.08   $\pm$   1.41    &       16.43 $\pm$ 1.65        &       15.01 $\pm$ 0.33      \\
327     &       12.08   $\pm$   1.21    &       12.52 $\pm$ 1.26        &       11.11 $\pm$ 0.16      \\
750     &       6.38    $\pm$   0.64    &       6.23 $\pm$ 0.63         &       5.25 $\pm$ 0.12      \\
966     &       5.09    $\pm$   0.26    &       4.97 $\pm$ 0.50         &       4.16 $\pm$ 0.08      \\
1400    &       3.56    $\pm$   0.18    &       3.55 $\pm$ 0.36         &       2.96 $\pm$ 0.05      \\
2700    &       1.89    $\pm$   0.10    &       1.89 $\pm$ 0.19         &       1.60 $\pm$ 0.03      \\
4850    &       0.98    $\pm$   0.05    &       1.03 $\pm$ 0.10         &       0.92 $\pm$ 0.01      \\
10695   &       0.42    $\pm$   0.02    &       0.42 $\pm$ 0.04         &       0.43 $\pm$ 0.01      \\
14900   &       0.25    $\pm$   0.01    &       0.27 $\pm$ 0.03         &       0.31 $\pm$ 0.01      \\
\hline
$r$ &  & 4 & 3 \\
$\chi^{2}$ &  & 3.45 & 10.60 \\
& $\Lambda$=0.10 & ($p$=0.1) &  \\
\hline
\end{tabular}
\end{center}
\end{table}

\subsubsection{3C401}

3C401 is a rather small ($D=$67.5 kpc) and low-redshifted ($z=0.201$) FR II-type radio source previously studied by Laing et al. (1983).
It coincides with a Seyfert galaxy and is characterised by a radio core of medium brightness equal to 28.54 $\pm$ 0.03 mJy 
(Hardcastle et al. 1998) and high axial
ratio of the lobes. In this case, there is no statistical preference of the KDA\,EXT model fit over the KDA
model, with the $\Lambda$ value being almost seven times larger than the level of significance.
The spectra resulting from the predicted flux densities $S_{\rm MOD}\,(KDA)$ and
$S_{\rm MOD}\,(KDA\,EXT)$ are listed in Table~8, along with their corresponding goodness of the fit to $\chi^{2}$  data, which
is evidently lower in the case of the KDA\,EXT fit.

\begin{table}[h]
\caption{\small Model flux densities $S_{MOD}$ from the KDA and KDA\,EXT fitting for 3C401 and their goodness of fit to the observed data $S_{\nu_{0}}$.}
\footnotesize
\begin{center}
\begin{tabular}{@{}rrrr}
\hline
$\nu_{0}$ [MHz] & $S_{\nu_{0}}\pm\Delta$ [Jy] & $S_{MOD}$ [Jy] & $S_{MOD}$ [Jy] \\
         & & KDA\,EXT  & KDA\\
\hline
\hline
74      &       31.38   $\pm$   3.14    &       42.52 $\pm$ 5.46       &        54.15 $\pm$ 4.29 \\
150     &       28.02   $\pm$   2.80    &       27.74 $\pm$ 2.65        &       30.32 $\pm$ 2.61 \\ 
178     &       23.55   $\pm$   2.36    &       24.52 $\pm$ 2.13        &       26.28 $\pm$ 2.80 \\
232     &       22.31   $\pm$   2.23    &       20.49 $\pm$ 1.60        &       21.07 $\pm$ 2.47 \\
325     &       15.31   $\pm$   1.53    &       16.17 $\pm$ 1.45        &       15.91 $\pm$ 1.63 \\
365     &       14.93   $\pm$   1.49    &       14.85 $\pm$ 0.80        &       14.41 $\pm$ 1.50 \\
740     &       8.66    $\pm$   0.87    &       8.64 $\pm$ 0.63         &       7.90 $\pm$ 0.78 \\
966     &       7.11    $\pm$   0.71    &       6.96 $\pm$ 0.46         &       6.28 $\pm$ 0.70 \\
1400    &       5.07    $\pm$   0.25    &       5.08 $\pm$ 0.26         &       4.55 $\pm$ 0.51 \\
2695    &       2.82    $\pm$   0.14    &       2.82 $\pm$ 0.15         &       2.54 $\pm$ 0.28 \\
4850    &       1.58    $\pm$   0.08    &       1.59 $\pm$ 0.15         &       1.48 $\pm$ 0.16 \\
4900    &       1.55    $\pm$   0.08    &       1.57 $\pm$ 0.14         &       1.47 $\pm$ 0.16 \\
5000    &       1.36    $\pm$   0.07    &       1.54 $\pm$ 0.14         &       1.44 $\pm$ 0.15 \\
8000    &       0.90    $\pm$   0.05    &       0.93 $\pm$ 0.09         &       0.93 $\pm$ 0.09 \\
10695   &       0.67    $\pm$   0.03    &       0.67 $\pm$ 0.07         &       0.70 $\pm$ 0.07 \\
10800   &       0.64    $\pm$   0.03    &       0.66 $\pm$ 0.07         &       0.69 $\pm$ 0.07 \\
14900   &       0.43    $\pm$   0.02    &       0.44 $\pm$ 0.05         &       0.51 $\pm$ 0.04 \\
\hline
$r$ &  & 4 & 3 \\
$\chi^{2}$ &  & 1.81 & 6.61 \\
& $\Lambda$=0.69 & ($p$=0.1) &   \\
\hline
\end{tabular}
\end{center}
\end{table}

\subsubsection{PKS 0235$-$19}

This is the only source in the sample not covered by the 3C catalogue due to its location in the southern hemisphere.
Nonetheless, PKS 0235$-$19 is listed in the Parkes Radio Catalogue (Wright \& Otrupcek 1990). Its nucleus has been classified as a
Seyfert 1 galaxy with broad Balmer lines (Tadhunter et al. 1998). The spectra predicted with the KDA and KDA\,EXT models are shown
in Figure 8. The relevant $\chi^{2}$ and $\Lambda$ values, along with the resulting flux densities calculated for these two models,
are presented in Table~9, confirming the preference for the KDA\,EXT solution.

\begin{table}[h]
\caption{\small Model flux densities $S_{MOD}$ from the KDA and KDA\,EXT fitting for PKS 0235$-$19 and their goodness of fit to the observed data $S_{\nu_{0}}$.}
\footnotesize
\begin{center}
\begin{tabular}{@{}rrrr}
\hline
$\nu_{0}$ [MHz] & $S_{\nu_{0}}\pm\Delta$ [Jy] & $S_{MOD}$ [Jy] & $S_{MOD}$ [Jy] \\
         & & KDA\,EXT  & KDA\\
\hline
\hline
74      &       35.66   $\pm$   3.57    &       45.45 $\pm$ 4.57        &       60.33 $\pm$ 6.31      \\
150     &       24.12   $\pm$   2.41    &       28.23 $\pm$ 2.84        &       32.30 $\pm$ 3.93      \\ 
160     &       26.50   $\pm$   2.65    &       27.01 $\pm$ 2.73        &       30.73 $\pm$ 3.76      \\
365     &       15.60   $\pm$   1.56    &       14.54 $\pm$ 1.46        &       14.37 $\pm$ 2.06      \\
408     &       13.34   $\pm$   1.33    &       13.29 $\pm$ 1.34        &       12.96 $\pm$ 1.89      \\
635     &       9.49    $\pm$   0.95    &       9.22 $\pm$ 0.93        &        8.50 $\pm$ 1.34      \\
1400    &       4.62    $\pm$   0.23    &       4.55 $\pm$ 0.46        &        3.94 $\pm$ 0.70      \\
2300    &       2.96    $\pm$   0.15    &       2.81 $\pm$ 0.28         &       2.41 $\pm$ 0.46      \\
2700    &       2.41    $\pm$   0.12    &       2.39 $\pm$ 0.24         &       2.05 $\pm$ 0.40      \\
3900    &       1.58    $\pm$   0.08    &       1.62 $\pm$ 0.16         &       1.42 $\pm$ 0.29      \\
4850    &       1.31    $\pm$   0.07    &       1.27 $\pm$ 0.13         &       1.14 $\pm$ 0.24      \\
10695   &       0.44    $\pm$   0.02    &       0.47 $\pm$ 0.05         &       0.51 $\pm$ 0.12      \\
11200   &       0.37    $\pm$   0.02    &       0.44 $\pm$ 0.04         &       0.49 $\pm$ 0.11      \\
\hline
$r$ &  & 4 & 3 \\
$\chi^{2}$ &  & 3.71 & 17.59 \\
& $\Lambda$=0.013 & ($p$=0.1) &  \\
\hline
\end{tabular}
\end{center}
\end{table}

\clearpage

\begin{figure}[h]
\includegraphics[width=8.5cm, height=8.5cm]{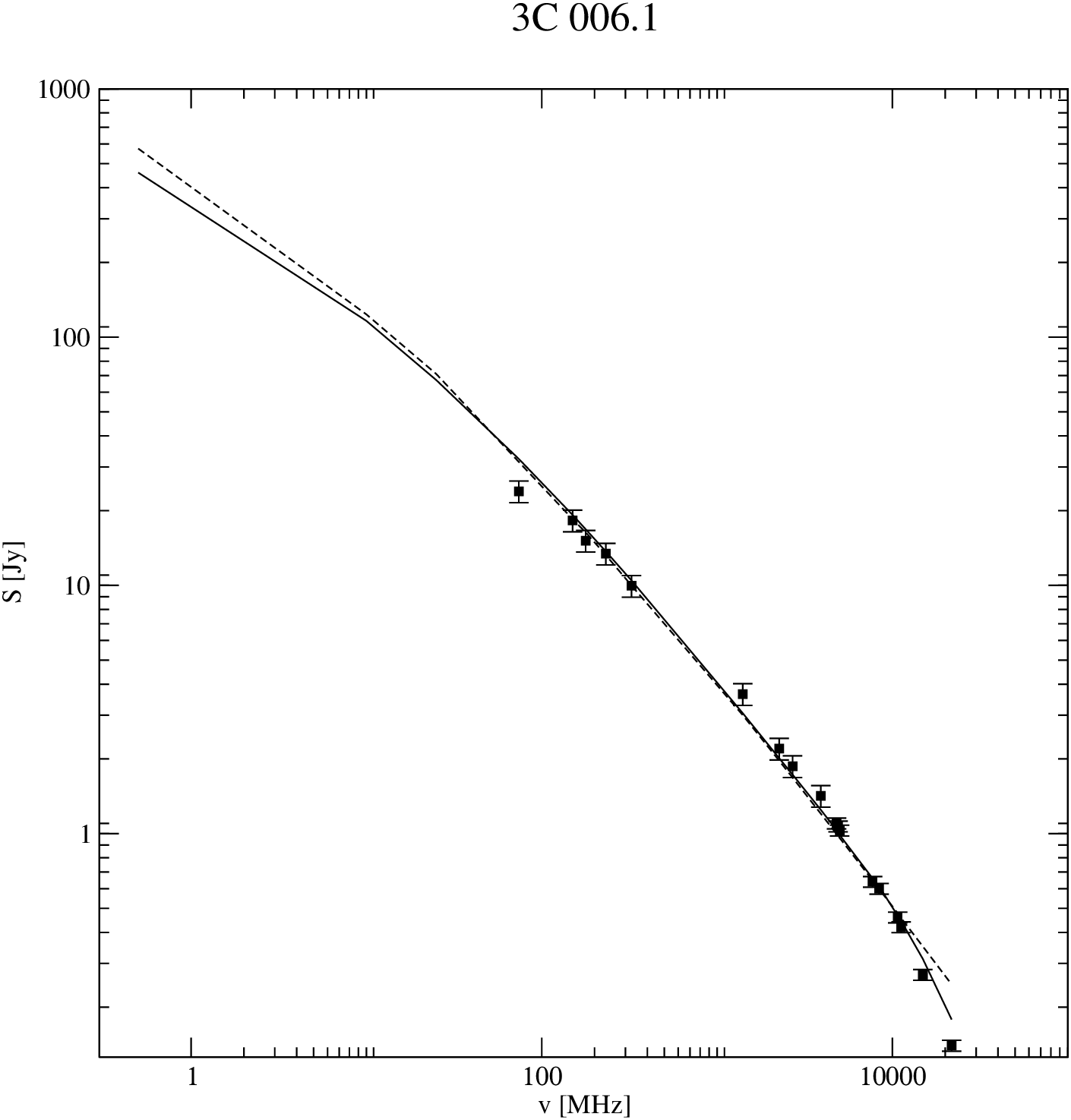}
\caption{Best KDA\,EXT fit (solid line) and KDA fit (dotted line) for radio galaxy 3C006.1. The values of flux density and 
frequency are presented in logarithmic scale. The observed flux densities and their errors are marked with data points.}
\label{1}
\end{figure}

\begin{figure}[h]
\includegraphics[width=8.5cm, height=8.5cm]{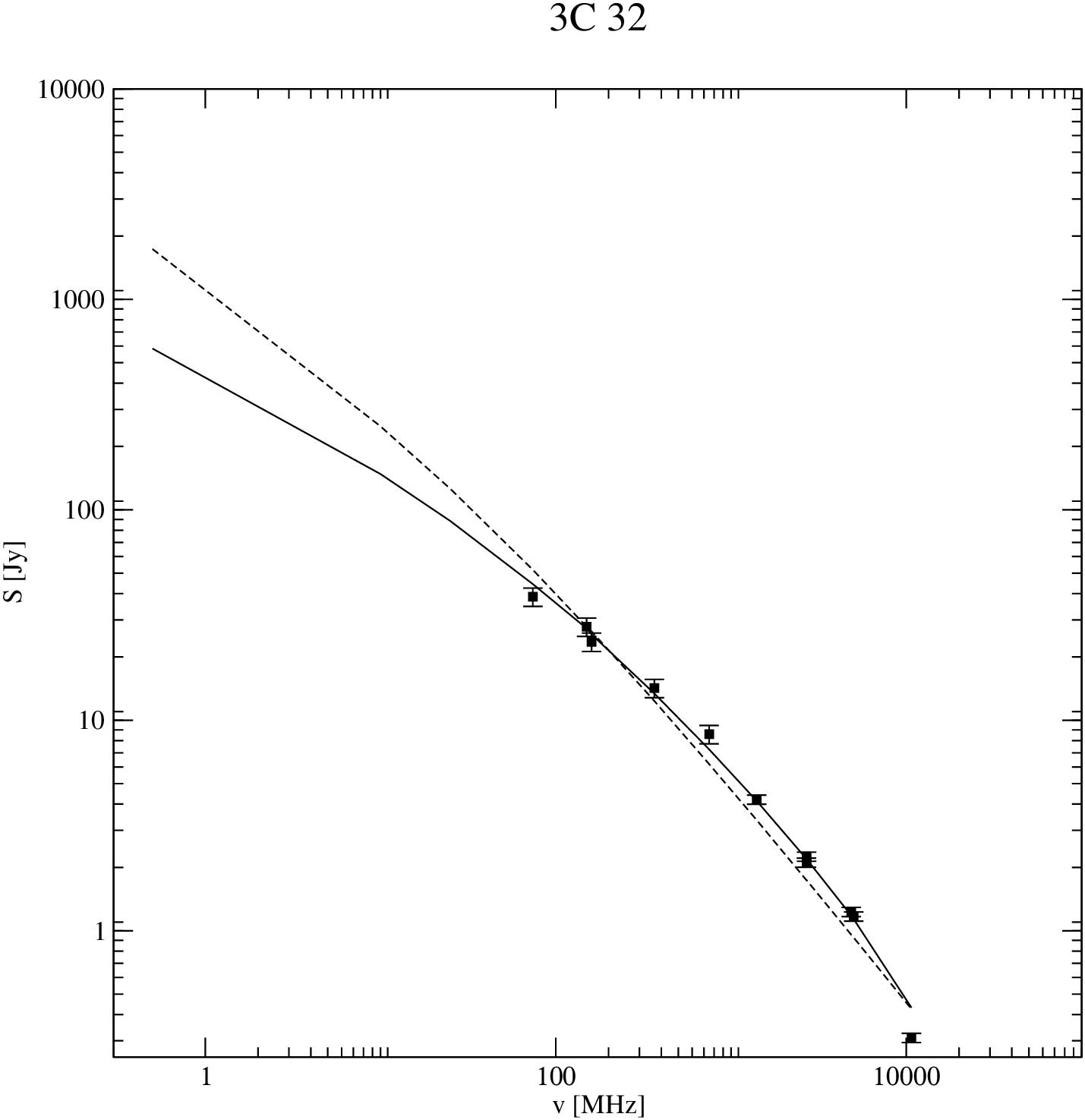}
\caption{As in Figure 1, but for 3C 32.}
\label{2}
\end{figure}

\begin{figure}[h]
\includegraphics[width=8.5cm, height=8.5cm]{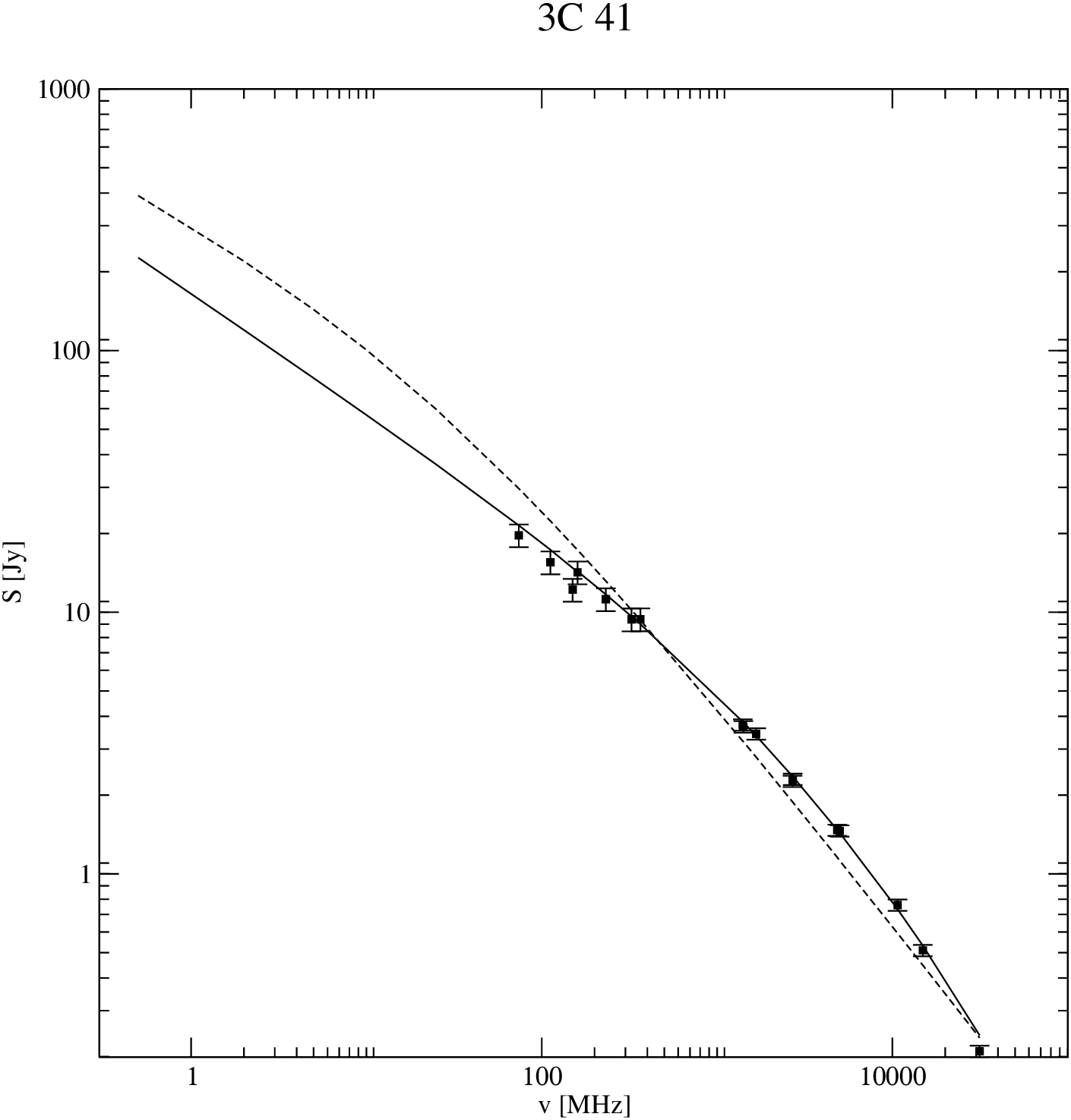}
\caption{As in Figure 1, but for 3C 41.}
\vspace{1.5cm}
\label{3}
\end{figure}

\begin{figure}[h]
\includegraphics[width=8.5cm, height=8.5cm]{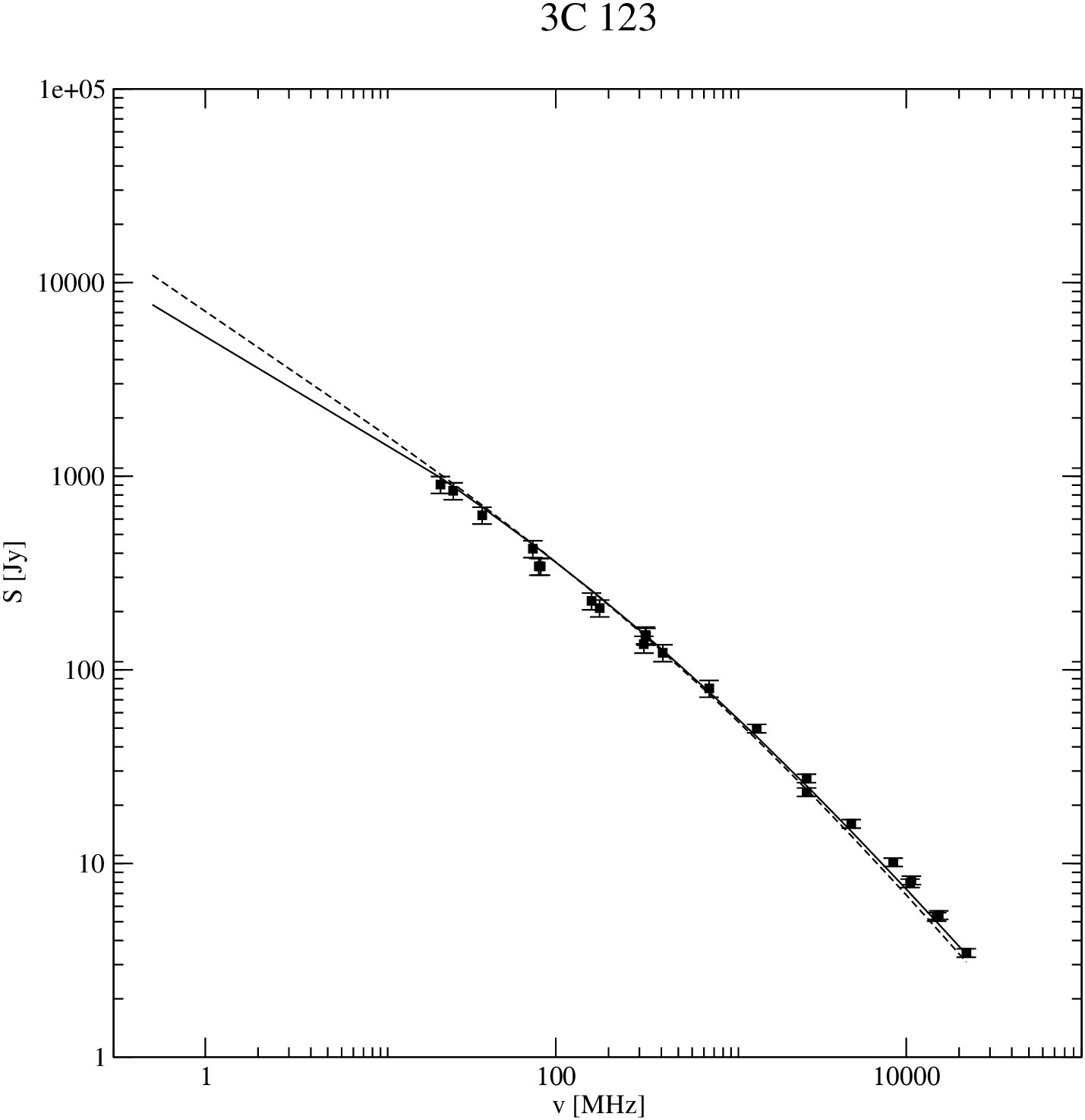}
\caption{As in Figure 1, but for 3C 123.}
\label{4}
\end{figure}

\begin{figure}[h]
\includegraphics[width=8.5cm, height=8.5cm]{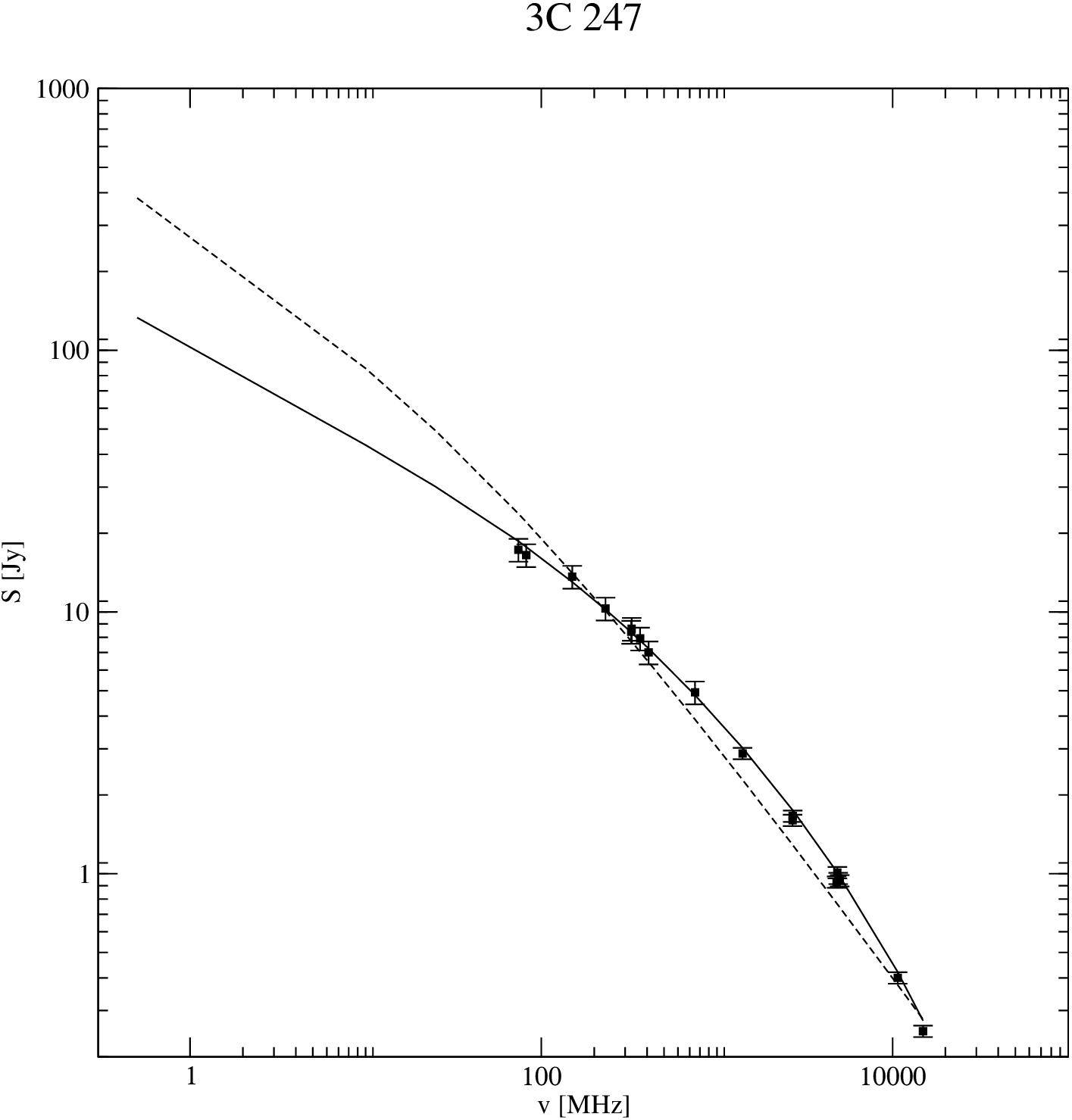}
\caption{As in Figure 1, but for 3C 247.}
\label{5}
\end{figure}

\begin{figure}[h]
\includegraphics[width=8.5cm, height=8.5cm]{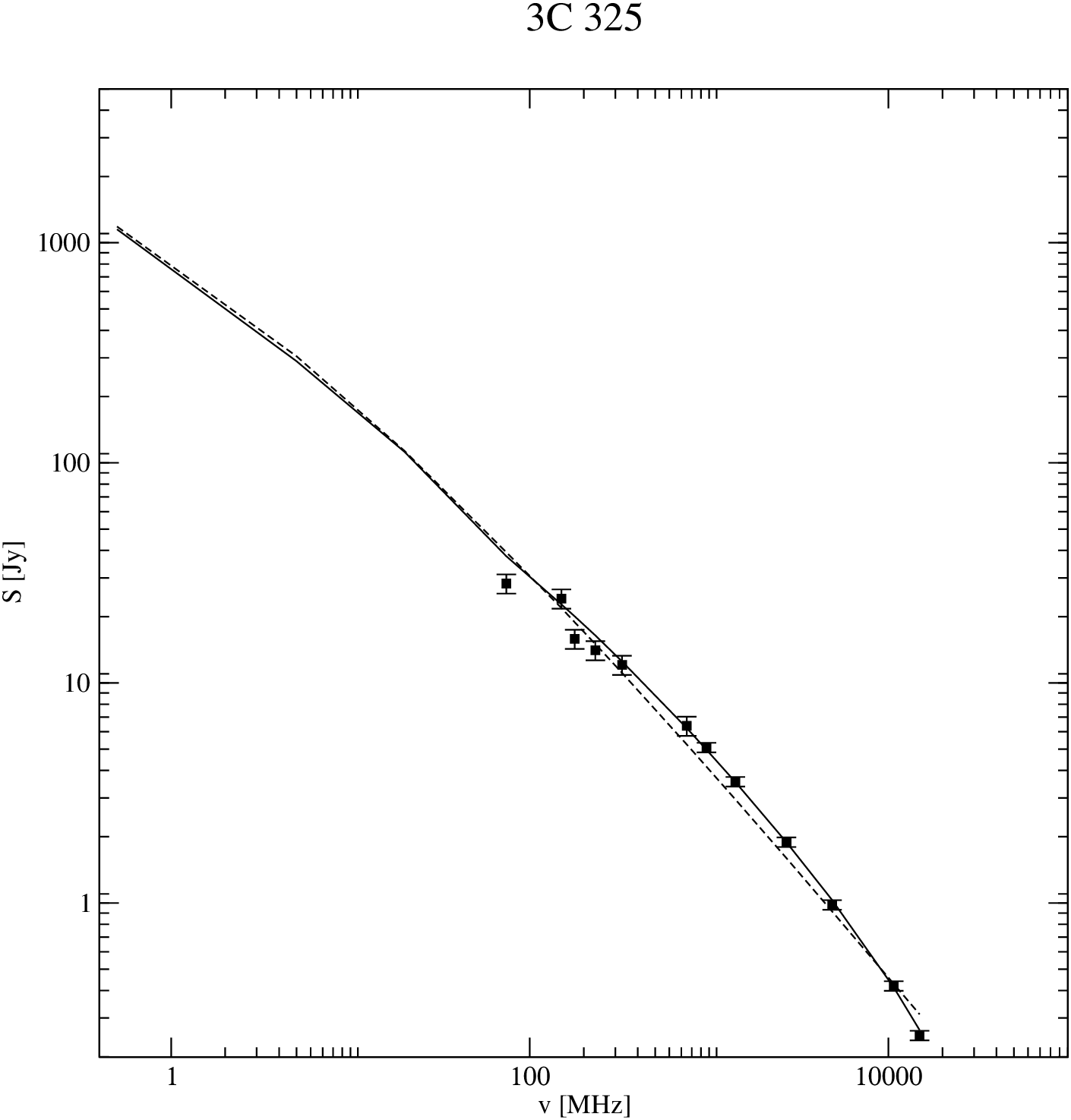}
\caption{As in Figure 1, but for 3C 325.}
\label{6}
\end{figure}

\begin{figure}[h]
\includegraphics[width=8.5cm, height=8.5cm]{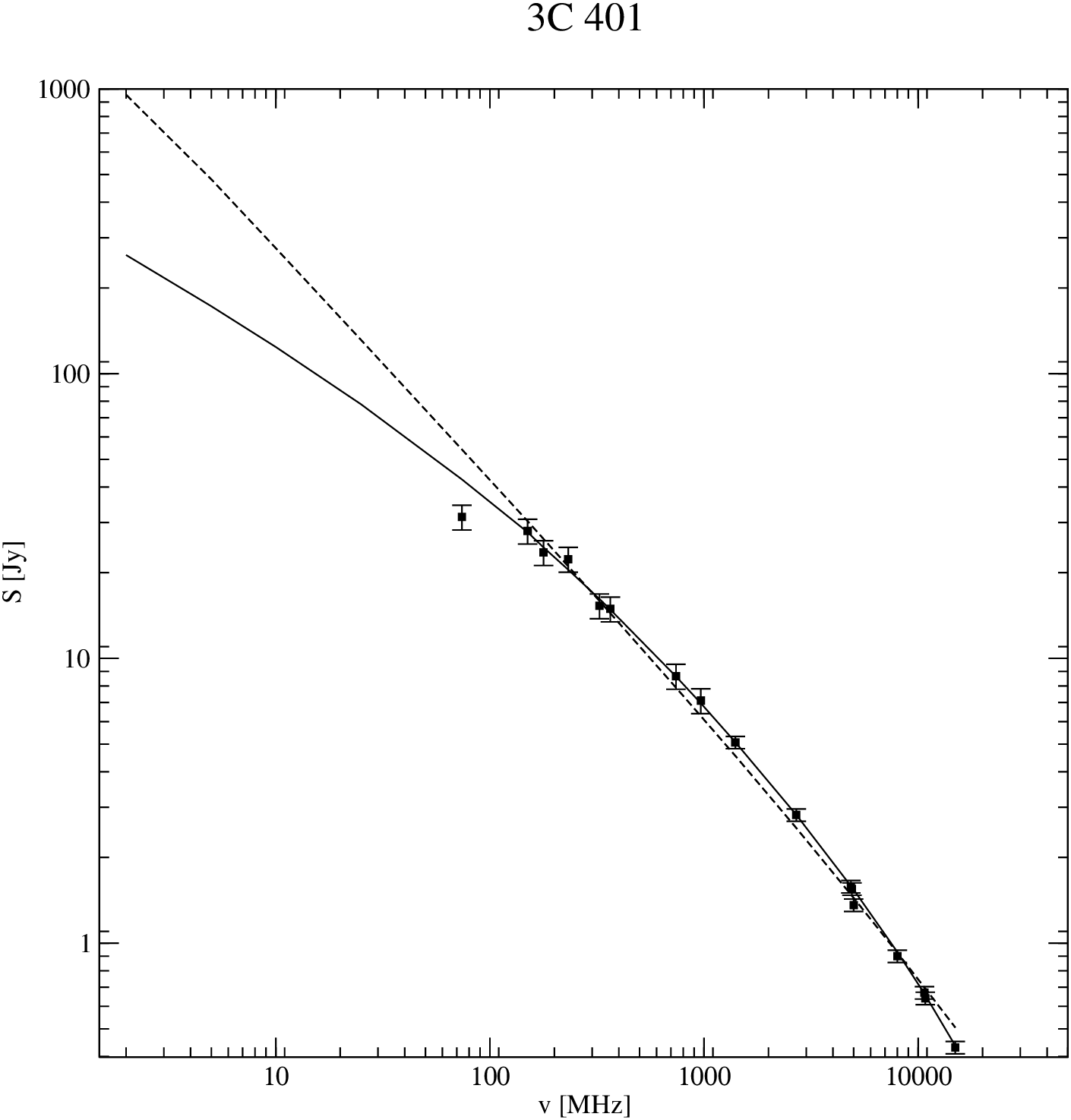}
\caption{As in Figure 1, but for 3C 401.}
\label{7}
\end{figure}

\begin{figure}[h]
\includegraphics[width=8.5cm, height=8.5cm]{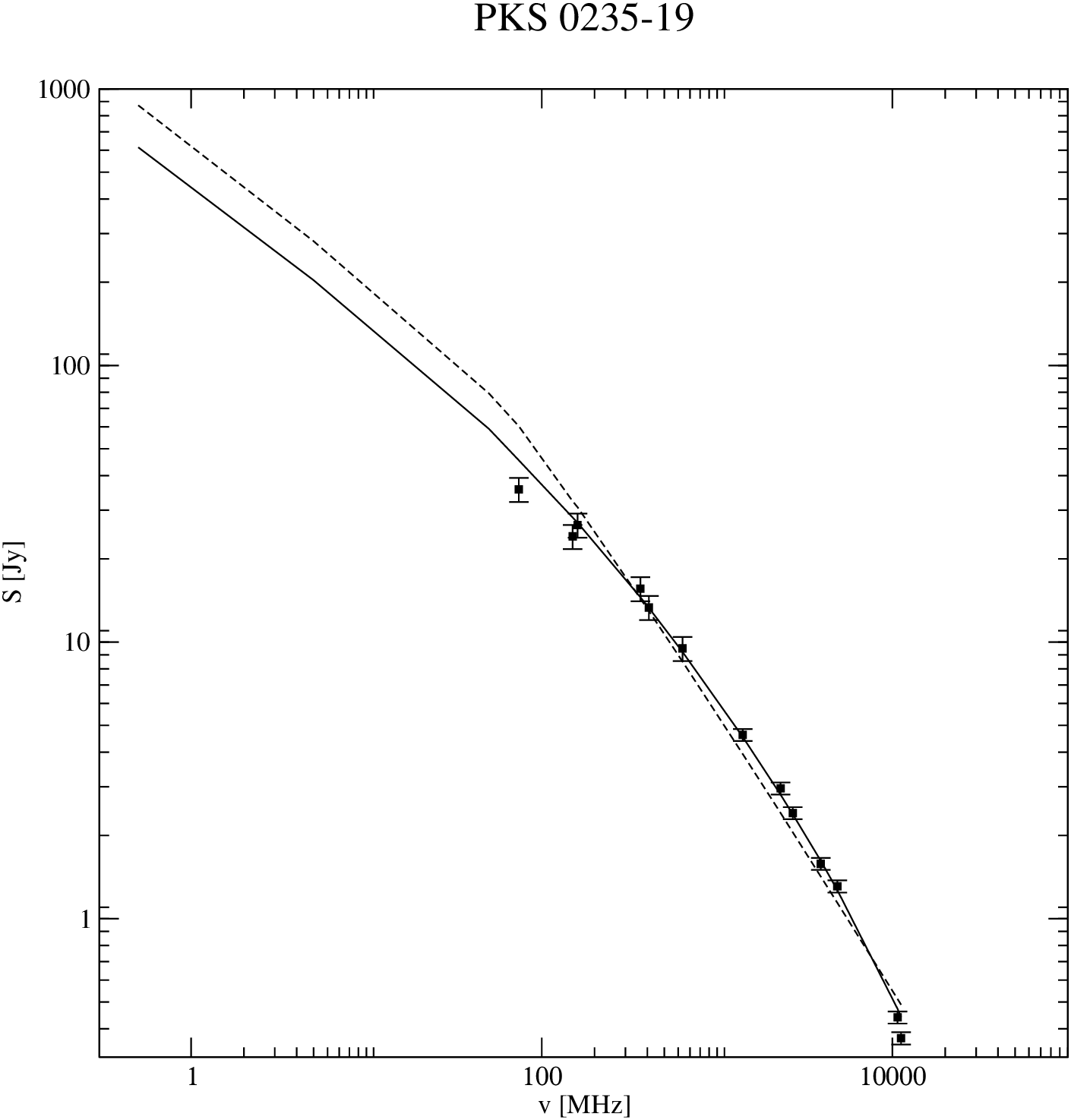}
\caption{As in Figure 1, but for PKS 0235$-$19.}
\label{8}
\end{figure}

\clearpage
\begin{table*}
\caption{Physical parameters of the sources from the sample derived from the best KDA fits for the entire available
radio spectrum (upper lines) and KDA\,EXT fits (bottom lines). Ages are given in Myr. The accuracy of their numerical 
determination is of the order of $\pm$ 0.01 Myr. 
}
\label{table:2}
\begin{tabular}{@{ }lllllllll}
\hline \hline
Name & 3C006.1 & 3C32 & 3C41 & 3C123 & 3C247 & 3C325 & 3C401 & PKS 0235$-$19  \\
\hline
t$_{\rm KDA}$ [Myr] & 19.09 & 45.71 & 9.12 & 2.31 & 26.55 & 31.61 & 0.20  & 25.69\\
t$_{\rm KDA\,EXT}$ [Myr] & 19.09 & 38.02 & 2.00 & 1.91 & 9.15 & 17.54  & 6.27 & 25.59  \\
t$_{\rm br}$ [Myr] & 18.43 & 35.91 & 1.73 & 1.82 & 7.31 & 16.60  & 5.71 & 23.22\\
\hline
$\alpha_{\rm inj}$ & 0.38 & 0.56 & 0.37 & 0.59 & 0.44 & 0.44 & 0.80 & 0.46  \\
& 0.35 & 0.41 & 0.46 & 0.54 & 0.38 & 0.46 & 0.43 & 0.46\\
$Q_{\rm jet}$ [W] & 17.1$\times$10$^{38}$ & 3.6$\times$10$^{38}$ & 1.9$\times$10$^{39}$ & 1.8$\times$10$^{39}$ &
4.1$\times$10$^{39}$ & 1.5$\times$10$^{39}$  & 3.9$\times$10$^{39}$ & 8.9$\times$10$^{38}$  \\
& 22.7$\times$10$^{38}$ & 4.9$\times$10$^{38}$ & 3.8$\times$10$^{39}$ & 1.5$\times$10$^{39}$  & 1.3$\times$10$^{39}$ &
8.11$\times$10$^{38}$ & 1.01$\times$10$^{38}$  & 8.1$\times$10$^{38}$\\
$\rho_{0}$ [kg/m$^{3}$] & 4.8$\times$10$^{-22}$ & 6.0$\times$10$^{-22}$ & 4.8$\times$10$^{-22}$ & 1.7$\times$10$^{-23}$
 & 3.5$\times$10$^{-23}$ & 3.5$\times$10$^{-22}$ & 1.8$\times$10$^{-25}$ &  2.5$\times$10$^{-23}$  \\
& 6.0$\times$10$^{-22}$ & 5.7$\times$10$^{-22}$ & 6.9$\times$10$^{-24}$ & 1.9$\times$10$^{-23}$ & 7.4$\times$10$^{-24}$ &
2.9$\times$10$^{-23}$ & 7.7$\times$10$^{-23}$  & 2.1$\times$10$^{-23}$ \\
$p_{\rm c}$ [N/m$^{2}$] & 2.5$\times$10$^{-11}$ & 3.9$\times$10$^{-12}$ & 5.1$\times$10$^{-11}$ & 3.1$\times$10$^{-11}$ &
7.1$\times$10$^{-12}$ & 1.91$\times$10$^{-14}$ &  4.72$\times$10$^{-11}$ & 3.01$\times$10$^{-12}$\\
& 3.1$\times$10$^{-10}$ & 5.1$\times$10$^{-12}$ & 1.6$\times$10$^{-11}$ & 4.3$\times$10$^{-11}$ & 8.0$\times$10$^{-12}$ &
5.11$\times$10$^{-12}$ & 2.19$\times$10$^{-11}$ & 2.51$\times$10$^{-12}$ \\
$U_{\rm c}$ [J] & 1.03$\times$10$^{54}$ & 5.16$\times$10$^{53}$ & 5.38$\times$10$^{53}$ & 1.32$\times$10$^{53}$ &
3.47$\times$10$^{53}$ & 1.50 $\times$10$^{54}$ & 2.45$\times$10$^{52}$ & 7.30$\times$10$^{53}$   \\
& 1.36$\times$10$^{54}$  & 5.87$\times$10$^{52}$ & 2.40 $\times$10$^{53}$ & 9.22$\times$10$^{52}$ & 3.01$\times$10$^{53}$ &
5.26$\times$10$^{53}$ & 2.0$\times$10$^{52}$ & 6.5$\times$10$^{53}$ \\
$B_{\rm eq}$ [nT] & 6.33 & 2.52 & 9.12 & 7.06 & 3.39 & 5.56 & 8.73 & 2.21\\
& 7.01 & 2.88 & 5.06 & 8.37 & 3.61 & 2.87 & 5.94 & 2.01\\
$v_{\rm h}/c$ & 0.018 & 0.012 & 0.026 & 0.074 & 0.01 & 0.015 & 0.38 & 0.017 \\
& 0.017 & 0.014 & 0.013 & 0.071 & 0.033 & 0.019 & 0.015 & 0.018 \\
\hline
\end{tabular}
\end{table*}

\section{Discussion}
~
This work concerns the KDA\,EXT model describing the dynamics of the
FR-II-type radio sources after recent cessation of their jet activity (i.e. having low $(t-t_{br})/t_{br}$ ratios 
not exceeding 0.25). KDA\,EXT is based on the KDA model and assumes
self-similarity of the particle flow. Thus one may expect this model not to be applicable in the
case of presumably very old radio sources whose jets terminated in the distant past.
Theoretical considerations along with some earlier modelling results suggest, rather, that, for
so-called radio relics (Murgia et al. 2011) in which aging of the radio structure is very evident and the self-similarity is
manifestly no longer valid, neither the KDA nor the KDA\,EXT models can correctly predict their observed radio spectra.
Following this supposition, the presented research was focused on those sources showing only modest spectral aging.

This paper is an extended illustration of the general predictions and qualitative analysis of the KDA\,EXT model, but is restricted due to the very 
limited sample of the analysed sources. The number of sources showing both the FR\,II structure and strongly curved radio spectra is 
small. Future work focussed on proper measurement data for a larger sample of such sources may help greatly in both constraining the parameters 
of the examined model and evaluation of its reliability.

The new results generally confirm those obtained earlier by Kuligowska (2017):
in the majority of the analysed cases, the solutions for individual sources summarised in Section\,4
still show that the KDA\,EXT model is useful and suitable for studying the sources with strongly bent radio spectra
in which central activity terminated relatively recently (or formally: with low ($t-t_{br}$)/$t_{br}$) ratios).
In the context of the presented results (All Figures and Table~10), the most controversial cases are radio sources 3C41 and
3C401. Figure~3 (3C41) shows that in the range of the observed flux densities for this source, the spectrum resulting from the KDA model is
initially much steeper than the one obtained with the KDA\,EXT model. However, the corresponding entries in Table~10 clearly indicate
that the $\alpha_{\rm inj}$ is much lower in the case of the KDA model. This difference is easy to explain if
only the wider range of the spectrum (adding the extremely low frequencies ranging from less than 1 MHz to 10 MHz) is presented. Thus the
spectra for both models have been calculated for lower (starting at 0.5 MHz) frequencies.
It should be noted that for the frequencies lower than 10 MHz there is no technical ability to observe radio flux
densities with the present-day ground-based telescopes, mostly due to the impact of the ionosphere. The analysis of this low-frequency model spectra 
confirms that the initial slope is effectively flatter in the case of the KDA\,EXT model. On the other hand, in the case of 3C401 the resulting KDA and KDA\,EXT plots
(Fig. 7) are not very different in the range of the lower observational frequencies, though the formal measurement uncertainties for both models are
rather large in this scope.
If, however, the KDA and KDA\,EXT models for this source are recalculated taking into account the extremely low frequencies lying
beyond the observed data points, it becomes clear that the discrimination between $\alpha_{\rm inj}$ values (also very evident in Table~10) 
and the overall difference between the two models are actually significant.

The source 3C123 is also an interesting case. Here, both model (KDA and KDA\,EXT) fits and their resulting $\chi^{2}$ values
are quite similar. Also the applied statistical analysis based on the likelihood ratio test does not confirm the preference of the
KDA\,EXT model. Once again, the solution presented for this source proves that very congruous model-predicted
spectral shapes can be obtained starting with not only different models, but also different values of their input parameters (compare
Table~10 and the paper of Brocksopp et al. (2011).

Regardless of the apparently curved spectra of the sample sources, an attempt was made to fit their observed spectra
with the classical KDA model with high values of the $\alpha_{\rm inj}$ parameter. Here,  conversely, it was assumed that these
source are still active and their jets still supply the radio structure. However, it is easy to notice that these resulting fits
do not reproduce the observed flux density properly and in the case of 3C401, they result in
a very young source age of 0.2 Myr.

In addition, the result of the likelihood ratio test clearly indicates that the difference between the KDA and KDA\,EXT models
calculated for this particular source is not statistically significant at the significance level p$=$0.10$\%$, despite the large
difference in their $\chi^{2}$ values. It may be concluded that there is no statistical preference for the KDA\,EXT, though the
subjective visual plot analysis favours this model. 

The results of the KDA\,EXT modelling presented in this work do not always reproduce the spectra of
the sample sources with satisfactory accuracy. There are numerous points departing from the model lines, especially
in the low-frequency parts of the predicted spectra. It should be noted that for some sample sources, a good fit of this
spectral range was actually achieved, while in other cases a number of low-frequency points lie far from the model (compare 3C325 or 3C401). The same
situation also applies (to a lesser extent) to the high-frequency parts of the spectra. 
Except for the underestimated values of the observed flux densities and their errors, the above inconsistency may result from the following factors:
\begin{enumerate}
\item Variation in the rate of particles transported in the jets (not included in the KDA\,EXT).
\item The source's restarting activity, resulting in another variation in the jet composition.
\item General inaccuracy in determining the initial KDA model parameters for the given source for their further use
in fitting the KDA\,EXT model.
\item Inaccuracy in estimating the geometrical parameters of the sources such as lobe length or axial ratio, possibly due to
the lack of sufficiently detailed radio maps.
\end{enumerate}

The KDA\,EXT model predicting the evolution of the lobes of FR\,II-type radio sources
after termination of the jet activity, tested here once again on a new sample of galaxies, provides a satisfactory solution
for at least four of the eight examined radio sources. Statistically, the KDA\,EXT model is at least as good, and often better, than the original KDA 
model, and as such it is a useful tool for fitting and studying the steepened spectra of FR\,II radio sources.

\noindent
\newline

\newpage

{\bf ACKNOWLEDGEMENTS}
\newline
\newline
\normalsize 

\hspace*{0.004cm} This research was supported by the Polish National Science Centre grant No. 2013/09/B/ST9/00599. 
The authors express sincere gratitude to the anonymous referee for the useful and essential comments on the manuscript. 

\hspace*{0.004cm}

{\bf REFERENCES}
\newline
\newline
Baars, J. W. M., Genzel, R., Pauliny-Toth, I. I. K., \& Witzel, A. 1977 {\it A\&A}, {\bf 61}, 99\\
Barai, P., \& Wiita, P.J. 2006 {\it MNRAS}, {\bf 372}, 381\\
Blandford, R. D., \& Rees M. J., 1974 {\it MNRAS} {\bf 169}, 395\\
Brocksopp, C., Kaiser, C.R., Schoenmakers, A.P. et al. 2011 {\it MNRAS}, {\bf 410}, 484\\
Colla et al. 1970 {\it A\& AS} {\bf 1}, 281\\ 
Chidi, E. Akujor, E., Ludke, I. W. A. et al. 1994 {\it A\&AS} {\bf 105}, 247\\
Cohen, A.S. et al., 2007 {\it AJ}, {\bf 134}, 1245\\
Edge, D. O., Shakeshaft, J. R., McAdam, W. B. et al. 1959. {Mem. R. Astron. Soc.}, {\bf 68}, {37} \\
Fanaroff, B. L., \& Riley, J. M. 1974 {\it MNRAS}, {\bf 167}, 31 \\
Giovannini, G., Feretti, L., Gregorini, L., \& Parma, P. 1988, {\it A\&A}, {\bf 199}, 73 \\
Hardcastle, M.J., Alexander, P., Pooley, G.G. et al. 1998 {MNRAS}, {\bf 296}, 445 \\
Heavens, A. F. \& Drury, L. O'C. 1988 {\it MNRAS}, {\bf 235}, {997} \\
Hurley-Walker, N. et a. 2017 {\it MNRAS}, {\bf 464}, 1146\\
Hurley-Walker, N. 2017 \\
Helmboldt, J. F., Kassim, N. E., Cohen, A. S. et al. 2008 {\it ApJS}, {\bf 174}, 313\\
Intema, H. T., Jagannathan, P., Mooley, K.P. \& Frail, D. A. 2017 {\it A\&A} {\bf 598}, A78\\
Jaffe, W., Perola, G. 1973 {\it A\&A} {\bf 26}, 423\\
Kaiser, C. R., 2000 {\it A\& A}, {\bf 362}, 447\\
Kaiser, C. R., \& Alexander P. 1997 {\it MNRAS}, {\bf 286}, 215\\
Kaiser, C. R., \& Cotter G. 2002 {\it MNRAS}, {\bf 336}, 649\\
Kaiser, C. R., Dennett-Thorpe, A., \& Alexander, P. 1997 {\it MNRAS}, {\bf 292}, 723\\
Kaiser, C. R., Schoenmakers, A. P., \& Rottgering, H. J. A. 2000 {\it MNRAS}, {\bf 315}, 381\\
King, I. R. 1972 {\it ApJ}, {\bf 174}, 123\\
Kuligowska E. 2017 {\it A \& A}, {\bf 598}, 93\\
Laing, R. A., Riley, J. M., Longair, M. S., 1983 {\it MNRAS}, {\bf 204}, 151\\
Looney, L. W., Hardcastle, M. J., 2000 {\it ApJ}, {\bf 534}, 172\\
Machalski, J., Chy\.zy, K. T., Stawarz, \L{}. et al. 2007 {\it A\&A} {\bf}, 43M\\ 
Manolakou, K., \& Kirk J. G., 2002 {\it A\&A} {\bf 391}, 127\\
Morganti, R., Killeen, N. E. B., Tadhunter, C. N. 1993 {\it MNRAS}, {\bf 263}, 1023\\
Neff, S. G., Roberts, L., \& Hutchings, J. B. 1995 {\it ApJS}, {\bf 99}, 349  \\
Pacholczyk, A. G., 1970 {\it "Series of Books in Astronomy and Astrophysics, San Francisco: Freeman, 1970"}\\
Rawlings, S., Eales, S., Lacy, M. 2001 {MNRAS}, {\bf 322}, 523 \\
Saikia, D. J., Jamrozy, M., Konar, C. et al. 2010 {\it Proceedings of the 25th Texas Symposium on Relativistic Astrophysics. December 6-10, 2010. Heidelberg} {\bf 14}\\
Shulevski, A., Morganti, R., Barthel, P. D. et al. 2015 {\it A\&A}, {\bf 583} 89\\
Spinrad, H., Marr, J., Aguilar, L., Djorgovski, S., 1985 {\it PASP}, {\bf 97}, 932 \\
Tadhunter, C. N., Morganti, R., Robinson, A., Dickson, R., Villar-Martin, M. et al. 1998 {\it MNRAS}, {\bf 298}, 1035 \\  
Véron-Cetty, M. P., Véron, P., 2006 {\it A\&A}, {\bf 455}, 773\\
Willot, C., Rawlings, S., \& Blundell, K. M. 2001 {\it MNRAS} {\bf 324}, 1 \\
Wright, E. \& Otrupcek R., 1990 {\it PASAu}, {\bf 8} 261 \\
\end{document}